\documentclass[preprint,12pt]{article}
\pdfoutput=1
\usepackage[utf8x]{inputenc}      
\usepackage{hyperref}
\usepackage{amsmath,amssymb,mathtools}
\usepackage[T1]{fontenc}          
\usepackage{booktabs,tabularx}
\usepackage{graphicx,subfig}
\usepackage{xspace}
\usepackage[usenames]{xcolor}\definecolor{fscolor}{RGB}{44,118,255}
\usepackage{geometry}
\usepackage{todonotes}
\usepackage{listings}
\usepackage[absolute]{textpos}
\usepackage[many]{tcolorbox}
\usepackage{xparse}
\usepackage[font=small,labelfont=bf,format=plain,margin=0.05\textwidth]{caption}
\usepackage{bbm}
\usepackage{tabularx}
\usepackage{cite}

\allowdisplaybreaks

\evensidemargin \oddsidemargin
\newcommand{\sbe}{s_\beta}
\newcommand{\cbe}{c_\beta}
\newcommand{\tbe}{t_\beta}
\newcommand{\sbb}{s_\beta^2}
\newcommand{\cbb}{c_\beta^2}
\newcommand{\tbb}{t_\beta^2}

\newcommand{\sbne}{s_{\beta_n}}
\newcommand{\cbne}{c_{\beta_n}}
\newcommand{\tbne}{t_{\beta_n}}
\newcommand{\sbnbn}{s_{\beta_n}^2}
\newcommand{\cbnbn}{c_{\beta_n}^2}

\newcommand{\sbce}{s_{\beta_c}}
\newcommand{\cbce}{c_{\beta_c}}
\newcommand{\tbce}{t_{\beta_c}}
\newcommand{\sbcbc}{s_{\beta_c}^2}
\newcommand{\cbcbc}{c_{\beta_c}^2}

\newcommand{\sae}{s_\alpha}
\newcommand{\cae}{c_\alpha}
\newcommand{\tae}{t_\alpha}
\newcommand{\saa}{s_\alpha^2}
\newcommand{\caa}{c_\alpha^2}

\newcommand{\DR}{{\ensuremath{\overline{\text{DR}}}}\xspace}

\newcommand{\OS}{{\text{OS}}\xspace}

\newcommand{\SUSY}{{\text{SUSY}}}

\newcommand{\heavy}{{\text{heavy}}}
\newcommand{\light}{{\text{light}}}

\newcommand{\FH}{\mbox{{\tt FeynHiggs}}\xspace}

\newcommand{\Fig}[1]{Fig.~\ref{#1}}
\newcommand{\Sec}[1]{Section~\ref{#1}}
\newcommand{\App}[1]{App.~\ref{#1}}
\newcommand{\Eq}[1]{Eq.~(\ref{#1})}

\newcommand{\EqsSub}[1]{Eqs.~(\ref{#1})}

\renewcommand{\Re}{\text{Re}}

\newcommand{\cp}{\ensuremath{{\cal CP}}}

\newcommand{\msusy}{M_\SUSY}

\newcommand{\tev}{\,\, \mathrm{TeV}}
\newcommand{\gev}{\,\, \mathrm{GeV}}

\newcommand{\order}[1]{\ensuremath{{\cal O}(#1)}}
\newcommand{\al}{\alpha}

\newcommand{\alt}{\al_t}
\newcommand{\alb}{\al_b}

\newcommand{\dL}[1]{\delta^{(#1)}}
\newcommand{\dOL}{\dL{1}}
\newcommand{\dTL}{\dL{2}}
\newcommand{\DTL}{\Delta^{(2)}}

\newcommand{\dZOL}[1]{\dOL Z_{#1}}
\newcommand{\dZTL}[1]{\dTL Z_{#1}}
\newcommand{\DZ}[1]{\DTL Z_{#1}}
\newcommand{\gl}{\big|_\text{gl}}
\newcommand{\bgl}{\bigg|_\text{gl}}

\newcommand{\dmOL}[1]{\dOL m^2_{#1}}
\newcommand{\dmTL}[1]{\dTL m^2_{#1}}

\newcommand{\Hpm}{{H^\pm}}

\newcommand{\Gpm}{{G^\pm}}

\begin{document}

\thispagestyle{empty}
\def\thefootnote{\fnsymbol{footnote}}

\begin{flushright}
DESY 18-188\\
MPP-2018-240
\end{flushright}
\vspace{3em}
\begin{center}
{\Large\bf Pole mass determination in presence of heavy particles}
\\
\vspace{3em}
{
Henning Bahl$^{a,b}$\footnote{email: henning.bahl@desy.de}
}\\[2em]
{\sl${}^a$DESY, Notkestra{\ss}e 85, D-22607 Hamburg, Germany}\\
{\sl${}^b$Max-Planck Institut f\"ur Physik, F\"ohringer Ring 6, D-80805 M\"unchen, Germany}
\def\thefootnote{\arabic{footnote}}
\setcounter{page}{0}
\setcounter{footnote}{0}
\end{center}
\vspace{2ex}
\begin{abstract}
{}

We investigate the determination of the Higgs-boson propagator poles in the MSSM. Based upon earlier works, we point out that in case of a large hierarchy between the electroweak scale and one or more SUSY masses a numerical determination with \DR Higgs field renormalization induces higher order terms which would cancel in a more complete calculation. The origin of these terms is the momentum dependence of contributions involving at least one of the heavy particles. We present two different methods to avoid their appearance. In the first approach, the poles are determined by expanding around the one-loop solutions. In the second approach, a ``heavy-OS'' Higgs field renormalization is employed in order to absorb the momentum dependence of heavy contributions. We will find that the first approach leads to an unwanted behavior of the Higgs boson mass predictions close to crossing points where two Higgs bosons that mix with each other are almost mass degenerate. These problems are avoided in the second approach, which became the default approach used in the public code \FH. Despite the discussion being very specific to the MSSM, the argumentation and the methods presented in this work are straightforwardly applicable to the determination of propagator poles in other models involving a large mass hierarchy.

\end{abstract}

\newpage
\tableofcontents
\newpage
\def\thefootnote{\arabic{footnote}}


\section{Introduction}
\label{sec:01_intro}

The mass measurement of the Higgs boson discovered at the Large Hadron Collider (LHC) by the ATLAS and CMS experiments fixed the last free parameter of the Standard Model (SM)~\cite{Aad:2012tfa,Chatrchyan:2012xdj,Aad:2015zhl}. It is, however, well known that beyond the SM (BSM) physics is needed to tackle several unresolved problems of both experimental and theoretical nature. One of the most common models for BSM physics is the Minimal Supersymmetric Standard Model (MSSM)~\cite{Nilles:1983ge,Haber:1984rc}, based upon the concept of supersymmetry (SUSY). Apart from adding a superpartner to each SM degree of freedom, it also introduces a second Higgs doublet leading to five physical Higgs bosons: At the tree-level, these are the \cp-even $h$ and $H$ boson, the \cp-odd $A$ boson and the charged $\Hpm$ bosons.

One of the MSSM's unique features is the fact that the Higgs sector is highly predictive. Using one of the non-SM-like Higgs masses as an input, the masses of the remaining Higgs bosons, of which one plays the role of the Higgs boson discovered at the LHC, are calculable in terms of only a few relevant parameters. In consequence, the mass of the Higgs boson discovered at the LHC can be used as a precision observable to constrain the MSSM parameter space. In order to profit from the high experimental precision, it is crucial to take radiative corrections into account. In the most direct approach, corrections to the Higgs-boson propagators are evaluated by the means of a fixed-order diagrammatic calculation (see~\cite{Borowka:2015ura,Goodsell:2016udb,Passehr:2017ufr,Harlander:2017kuc,Borowka:2018anu} for recent works and references therein). In this approach, large logarithms appear in the calculation if one or more of the SUSY particles become much heavier than the electroweak scale. These can spoil the convergence of the perturbative expansion. Therefore, effective field theory (EFT) techniques are used to resum these logarithms (see~\cite{Vega:2015fna,Lee:2015uza,Bagnaschi:2017xid,Bahl:2018jom,Harlander:2018yhj} for recent works and references therein). Since typically no higher dimensional operators are included in the effective Lagrangian,\footnote{See \cite{Bagnaschi:2017xid} for a study taking higher dimensional operators into account.} EFT calculations miss terms which would be suppressed in case of a heavy SUSY scale. To obtain a precise predictions for low, intermediary and high SUSY scales, hybrid calculations combining the two approaches have been developed \cite{Hahn:2013ria,Bahl:2016brp,Athron:2016fuq,Staub:2017jnp,Bahl:2017aev,Athron:2017fvs,Bahl:2018jom}.

In this paper, we will focus on the hybrid approach which is implemented into the publicly available code \FH~\cite{Heinemeyer:1998yj,Heinemeyer:1998np,Hahn:2009zz,
Degrassi:2002fi,Frank:2006yh,Hahn:2013ria,Bahl:2016brp,Bahl:2017aev,Bahl:2018qog}. Its main idea is to add the logarithms resummed by means of an EFT calculation directly to the fixed-order result, namely the diagrammatic corrections to the Higgs-boson propagators. Appropriate subtraction terms ensure that double-counting of logarithms contained in both results is avoided.

The physical Higgs masses squared are then determined by finding the poles of the Higgs-boson propagator matrix as calculated in the fixed-order approach (supplemented by the logarithms resummed in the EFT approach) and taking their real part. This pole determination is the main topic of this paper. We discuss three different methods.

In the first method, used in \FH up to version \texttt{2.13.0}, the poles are determined by finding them numerically employing a \DR renormalization of the Higgs fields. As already found in \cite{Bahl:2017aev}, the momentum dependence of contributions involving heavy superpartners induces higher order terms which would cancel in a more complete calculation.

In \cite{Bahl:2017aev}, a solution for this issue was proposed in the limit of all non-SM-like Higgs bosons being heavy by determining the Higgs-boson propagator poles at a fixed order. As a second method, we will extend this approach to be applicable also for the case of light non-SM-like Higgs bosons. We will show, however, that the truncation at a fixed-order leads to unwanted discontinuities in the prediction of the Higgs boson masses close to crossing points, where two Higgs bosons that mix with each other are almost mass degenerate.

Therefore, we will present a third method not suffering from the problems of the other two methods. It resembles the first method since the poles are determined numerically. In contrast to the first method, the Higgs field renormalization is, however, not renormalized in the \DR scheme. Whereas the Higgs field renormalization drops out of the calculation at every complete order of the calculation, it is used at higher incomplete orders to cancel the momentum dependence of contributions involving heavy superpartners. We refer to this renormalization scheme as ``heavy-OS'' scheme. A similar renormalization scheme has already been introduced in \cite{Bahl:2018jom} in order to combine an EFT calculation suitable for light non-SM-like Higgs bosons with the fixed-order calculation implemented in \FH. In this paper, we extend this scheme to the case of scenarios involving complex \cp-violating parameters and apply it to the pole determination. This method is employed by default in \FH since version~\texttt{2.14.3}.  As we will show, its implementation was crucial for the precise prediction of the Higgs boson masses in some of the MSSM Higgs benchmark scenarios presented in~\cite{Bahl:2018zmf}. Moreover, we show that with this method a previously missed next-to-leading logarithmic contribution is taken correctly into account.

The paper is structured as follows. In \Sec{sec:02_MSSMintro}, we describe the MSSM Higgs sector and its renormalization. In \Sec{sec:03_p2heavy}, we shortly review the decoupling behavior of heavy contributions in general theories (without being specific to the MSSM). The consequences for the Higgs pole determination in the MSSM are then discussed in \Sec{sec:04_poledet}. Numerical results are presented in \Sec{sec:05_results}.


\section{The MSSM Higgs sector and its renormalization}
\label{sec:02_MSSMintro}

In this Section, we present a short overview of the MSSM Higgs sector and its renormalization. We restrict our discussion to the one- and two-loop level. At the two-loop level, we neglect the electroweak gauge couplings and assume vanishing external momentum. In addition, we neglect all first and second generation as well as the tau Yukawa coupling. This corresponds to the accuracy level of the fixed-order corrections implemented as default in \FH~\cite{
Chankowski:1992er,Dabelstein:1994hb,Pierce:1996zz,
Heinemeyer:1998yj,Heinemeyer:1998np,Degrassi:2001yf,
Brignole:2001jy,Brignole:2002bz,Degrassi:2002fi,Dedes:2003km,
Heinemeyer:2004xw,Frank:2006yh,Heinemeyer:2007aq,Hahn:2009zz,
Hollik:2014wea,Hollik:2014bua,Hollik:2015ema,Hahn:2015gaa}.\footnote{See~\cite{Borowka:2014wla,Borowka:2015ura,Borowka:2018anu} for studies partly including the dependence on the external momentum at the two-loop level in the \FH framework.} The discussion is especially focused on the Higgs field renormalization, which is of particular importance in this paper.

The conventions follow closely those of e.g.~\cite{Frank:2006yh,Hollik:2014bua}.


\subsection{Higgs sector at the tree level}

The Higgs sector of the MSSM consists out of two Higgs doublets $\mathcal{H}_1$ and $\mathcal{H}_2$ with opposite hypercharge. They are given by
\begin{align}\label{Eq:HiggsExpansion}
\mathcal{H}_1= \begin{pmatrix} v_1 + \frac{1}{\sqrt{2}}(\phi_1-i \chi_1)\\ -\phi_1^- \end{pmatrix},\hspace{2cm} \mathcal{H}_2= \begin{pmatrix} \phi_2^+\\ v_2 + \frac{1}{\sqrt{2}}(\phi_2+i \chi_2) \end{pmatrix},
\end{align}
where $v_1$ and $v_2$ are the vacuum expectation values (vevs) of the doublets. The ratio of the two vevs is denoted by
\begin{align}
\tan\beta = \frac{v_2}{v_1}.
\end{align}
The mass matrices of the component fields are diagonalized by the means of unitary transformations yielding the mass eigenstates. The transformations can be written as
\begin{align}\label{eq:HiggsMassStateTrafo}
\begin{pmatrix} h \\ H \\ A \\ G \end{pmatrix} = \textbf{U}_{n} \begin{pmatrix} \phi_1 \\ \phi_2 \\ \chi_1 \\ \chi_2 \end{pmatrix}, \hspace{0.5cm}
\begin{pmatrix} H^\pm \\ G^\pm \end{pmatrix} = \textbf{U}_{c} \begin{pmatrix} \phi_1^\pm \\ \phi_2^\pm \end{pmatrix}.
\end{align}
The unitary matrices $\textbf{U}_n$ and $\textbf{U}_c$ can be parametrized using the three mixing angles $\alpha$, $\beta_n$ and $\beta_c$,
\begin{align}
\textbf{U}_n = \begin{pmatrix}
-\sae & \cae & 0      & 0     \\
 \cae & \sae & 0      & 0     \\
0     & 0    & -\sbne & \cbne \\
0     & 0    &  \cbne & \sbne
\end{pmatrix}, \hspace{0.5cm}
\textbf{U}_c = \begin{pmatrix}
-\sbce & \cbce \\
\cbce & \sbce
\end{pmatrix}.
\end{align}
where we introduced the short-hand notation
\begin{align}
c_\gamma = \cos\gamma, \hspace{.3cm} s_\gamma = \sin\gamma, \hspace{.3cm} t_\gamma = \tan\gamma
\end{align}
for a generic angle $\gamma$ (the abbreviation for the tangent is used later).

By fixing
\begin{align}
\tan 2\alpha &= \frac{M_A^2+M_Z^2}{M_A^2-M_Z^2}\tan 2\beta,
\end{align}
where  $-\pi/2 \le \alpha \le 0$ is chosen by convention, and
\begin{align}
\beta_c &= \beta_n = \beta
\end{align}
the resulting mass matrices are diagonal. $M_Z$ is the mass of the $Z$~boson. $M_A$ denotes the mass of the $A$ boson, which is typically used as input parameter in the \cp-conserving case (i.e., if all parameters are real). In the \cp-violating case, the mass of the $\Hpm$ boson is used as input parameter.

The resulting tree-level masses are,
\begin{subequations}\label{eq:tree-level_masses}
\begin{align}
m_\Hpm^2  &= M_A^2 + M_W^2 \\
m_{h,H}^2 &= \frac{1}{2}\left(M_A^2 + M_Z^2 \mp \sqrt{(M_A^2 + M_Z^2)^2 - 4 M_A^2 M_Z^2 c_{2\beta}^2}\right), \\
m_G^2     &= m_\Gpm^2 = 0,
\end{align}
\end{subequations}
where $M_W$ is the mass of the $W$~boson.

The irreducible two-point vertex functions at lowest order, marked by the superscript ``0'', are diagonal and given by
\begin{subequations}
\begin{align}
\mathbf{\Gamma}^{(0)}_{hHAG}(p^2)     = i\left(p^2 \mathbf{1}_{4\times 4} - \mathbf{M}_{hHAG}\right),\\
\mathbf{\Gamma}^{(0)}_{\Hpm\Gpm}(p^2) = i\left(p^2 \mathbf{1}_{2\times 2} - \mathbf{M}_{\Hpm\Gpm}\right),
\end{align}
\end{subequations}
where $\mathbf{M}_{hHAG}$ and $\mathbf{M}_{\Hpm\Gpm}$ are the diagonal tree-level mass matrices with the entries given in \EqsSub{eq:tree-level_masses}.


\subsection{Renormalization of the Higgs sector}

At higher orders, loop corrections modify the simple tree-level expressions presented above,
\begin{subequations}
\begin{align}
i\left(p^2 \mathbf{1}_{4\times 4} - \mathbf{M}_{hHAG}\right) \rightarrow i\left(p^2 \mathbf{1}_{4\times 4} - \mathbf{M}_{hHAG} + \mathbf{\hat\Sigma}_{hHAG}(p^2)\right), \\
i\left(p^2 \mathbf{1}_{2\times 2} - \mathbf{M}_{\Hpm\Gpm}\right) \rightarrow i\left(p^2 \mathbf{1}_{2\times 2} - \mathbf{M}_{\Hpm\Gpm} + \mathbf{\hat\Sigma}_{\Hpm\Gpm}(p^2)\right),
\end{align}
\end{subequations}
where the matrices $\mathbf{\hat\Sigma}$ mark the renormalized loop corrections to the Higgs-boson propagators, called self-energies. Note that those matrices are not diagonal. Consequently, the loop corrections lead to mixing of the tree-level mass eigenstates.

For a description of the mass and tadpole renormalization, we refer to the discussions in e.g.~\cite{Frank:2006yh,Hollik:2014bua}. Here, we instead focus on the Higgs field renormalization.


\subsubsection{Higgs field renormalization}
\label{subsubsec:fieldren}

To render the results for physical observables UV finite, mass and tadpole renormalization is sufficient. If Green's functions involving external Higgs bosons are required to be UV finite, also the Higgs fields have to be renormalized.

We introduce the field renormalization in the original gauge eigenstate basis (up to the two-loop order),
\begin{align} \label{eq:introfieldren}
\begin{pmatrix}
\mathcal{H}_1 \\ \mathcal{H}_2
\end{pmatrix}
\rightarrow
\mathbf{Z_\mathcal{H}}
\begin{pmatrix}
\mathcal{H}_1 \\ \mathcal{H}_2
\end{pmatrix}
=
\begin{pmatrix}
1 + \frac{1}{2}\dZOL{11} + \frac{1}{2}\DZ{11} & \frac{1}{2}\dZOL{12} + \frac{1}{2}\DZ{12} \\
\frac{1}{2}\dZOL{21} + \frac{1}{2}\DZ{21} & 1 + \frac{1}{2}\dZOL{22} + \frac{1}{2}\DZ{22}
\end{pmatrix}
\begin{pmatrix}
\mathcal{H}_1 \\ \mathcal{H}_2
\end{pmatrix}
\end{align}
with
\begin{align}\label{eq:dZ_def}
\DZ{ij} = \dZTL{ij} - \frac{1}{4}\left(\dZOL{ij}\right)^2\hspace{.5cm} (i,j = 1,2).
\end{align}
We label the components of $\mathbf{Z_\mathcal{H}}$ by $Z_{ij}$ ($i,j=1,2$). Note that in contrast to e.g.~\cite{Hollik:2014bua}, we also introduced an off-diagonal field renormalization. This off-diagonal field renormalization is not required in order to cancel UV divergencies. We will, however, use this renormalization prescription in \Sec{sec:04_poledet} to prevent the introduction of unwanted higher-order terms through the determination of the Higgs-boson propagator poles. Note also that the field renormalization as introduced in \Eq{eq:introfieldren} goes beyond the one employed in \cite{Bahl:2018jom}: we do not assume $Z_{12} = Z_{21}$ here.

For simplification, we restrict us here to the case of an hermitian field renormalization matrix,
\begin{align}\label{eq:hermitianZH}
\mathbf{Z_\mathcal{H}} = \mathbf{Z_\mathcal{H}}^\dagger.
\end{align}
Consequently, $Z_{11}$ and $Z_{22}$ have to be real and $Z_{21}$ has to be the complex conjugated of $Z_{12}$.

The field renormalization constants in the mass eigenstate basis are obtained straightforwardly by the rotation introduced in \Eq{eq:HiggsMassStateTrafo},
\begin{align}
\begin{pmatrix}h \\ H \\ A \\ G \end{pmatrix} &\rightarrow
\mathbf{Z}_{hHAG}
\begin{pmatrix}h \\ H \\ A \\ G \end{pmatrix}, \hspace{.5cm}
\begin{pmatrix} H^\pm \\ G^\pm \end{pmatrix}\rightarrow
\mathbf{Z}_{H^\pm G^\pm}
\begin{pmatrix} H^\pm \\ G^\pm \end{pmatrix},
\end{align}
with
\begin{align}
\mathbf{Z}_{hHAG} &=
\begin{pmatrix}
1 + \frac{1}{2}\delta Z_{hh} &     \frac{1}{2}\delta Z_{hH} &     \frac{1}{2}\delta Z_{hA} &     \frac{1}{2}\delta Z_{hG} \\
    \frac{1}{2}\delta Z_{hH} & 1 + \frac{1}{2}\delta Z_{HH} &     \frac{1}{2}\delta Z_{HA} &     \frac{1}{2}\delta Z_{HG} \\
    \frac{1}{2}\delta Z_{hA} &     \frac{1}{2}\delta Z_{HA} & 1 + \frac{1}{2}\delta Z_{AA} &     \frac{1}{2}\delta Z_{AG} \\
    \frac{1}{2}\delta Z_{hG} &     \frac{1}{2}\delta Z_{HG} &     \frac{1}{2}\delta Z_{AG} & 1 + \frac{1}{2}\delta Z_{GG}
\end{pmatrix}, \label{eq:ZhHAG}\\
\mathbf{Z}_{H^\pm G^\pm} &=
\begin{pmatrix}
1 + \frac{1}{2}\delta Z_{H^\pm H^\pm} &     \frac{1}{2}\delta Z_{H^\pm G^\pm} \\
    \frac{1}{2}\delta Z_{G^\pm H^\pm} & 1 + \frac{1}{2}\delta Z_{G^\pm G^\pm}
\end{pmatrix},
\end{align}
where $\delta Z_{ij} = \dZOL{ij}+\dZTL{ij}$ ($i,j=h,H,A,G$). Explicit expressions for these one- and two-loop counterterms can be found in \App{app:07_ren}.

As a consequence of \Eq{eq:hermitianZH}, all field renormalization constants in the \cp-even subsector, involving $h$ and $H$, get contributions only from the real parts of $\mathbf{Z_\mathcal{H}}$. The same holds true for the $\cp$-odd subsector, involving $A$ and $G$, and the diagonal field renormalization constants in the charged Higgs sector. In contrast, the field renormalization constants connecting the $\cp$-even and the $\cp$-odd subsectors (e.g.\ $\delta Z_{hA}$) receive contributions only from the imaginary parts of $\mathbf{Z_\mathcal{H}}$. The off-diagonal field renormalization constants in the charged Higgs sector get contributions from both the real and imaginary parts of $\mathbf{Z_\mathcal{H}}$.


\subsubsection[Renormalization of \texorpdfstring{$\tan\beta$}{tanb}]{Renormalization of \texorpdfstring{$\boldsymbol{\tan}\boldsymbol{\beta}$}{tanb}}
\label{sec:tanb_ren}

The vevs $v_1$ and $v_2$ are renormalized according to
\begin{align}\label{eq:vevren}
\begin{pmatrix}
v_1 \\
v_2
\end{pmatrix}
\rightarrow&
\begin{pmatrix}
1 + \frac{1}{2}\dZOL{11} + \frac{1}{2}\DZ{11} & \frac{1}{2}\Re\left(\dZOL{12} + \DZ{12}\right) \\
\frac{1}{2}\Re\left(\dZOL{21} + \DZ{21}\right) & 1 + \frac{1}{2}\dZOL{22} + \frac{1}{2}\DZ{22}
\end{pmatrix}
\begin{pmatrix}
v_1 + \dOL v_1 + \dTL v_1 \\
v_2 + \dOL v_2 + \dTL v_2
\end{pmatrix}
\end{align}
The appearance of the field renormalizaton constants follows from the renormalization prescription in \Eq{eq:introfieldren}.

The difference of the one-loop vev counterterms is UV~finite~\cite{Chankowski:1992er,Dabelstein:1994hb}. Therefore, we set
\begin{align}\label{eq:dv1eqdv2OL}
\frac{\dOL v_1}{v_1} = \frac{\dOL v_2}{v_2}.
\end{align}
In the approximation of vanishing electroweak gauge couplings, which we apply at the two-loop level, also the difference of the two-loop vev counterterms is UV~finite~\cite{Sperling:2013xqa}. So, we set
\begin{align}\label{eq:dv1eqdv2TL}
\frac{\dTL v_1}{v_1}\bgl = \frac{\dTL v_2}{v_2}\bgl,
\end{align}
where the subscript ``gl'' is used to indicate the limit of vanishing electroweak gauge couplings, which is also often called gaugeless limit.

Applying these simplifications, the resulting one- and two-loop counterterms of $\tan\beta$, introduced via the renormalization transformation
\begin{align}
\tbe\rightarrow \tbe + \dOL\tbe + \dTL\tbe,
\end{align}
are then given by
\begin{subequations}
\begin{align}
\dOL\tbe ={}& \frac{1}{2}\tbe\left(\dZOL{22} - \dZOL{11}\right) + \frac{1}{4}\left(1-\tbb)(\dZOL{12}+\dZOL{21}\right), \\
\dTL\tbe\gl ={}&\bigg[
                 \frac{1}{2}\tbe\left(\dZTL{22}-\dZTL{11}\right) +\frac{1}{4}\left(1-\tbb\right)\left(\dZTL{12}+\dZTL{21}\right) \nonumber\\
&\hspace{.1cm} - \frac{1}{2}\tbe\left(\dZOL{22}-\dZOL{11}\right)\left(\frac{1}{4}\left(\dZOL{22}-\dZOL{11}\right)+\dZOL{11}\right) \nonumber\\
&\hspace{.1cm} - \frac{1}{8}\left((1-2\tbb)\dZOL{11} + \tbb\dZOL{22}\right)\left(\dZOL{12}+\dZOL{21}\right) \nonumber\\
&\hspace{.1cm} - \frac{1}{16}(1-\tbb)\left((1+\tbe)\left((\dZOL{12})^2 + (\dZOL{21})^2\right) + 2\tbe\dZOL{12}\dZOL{21}\right)
\bigg]_\text{gl}.
\end{align}
\end{subequations}


\subsubsection{Renormalization conditions for the Higgs fields}

The UV divergent part of the field renormalization constants is fixed by demanding that the derivatives of the Higgs self-energies with respect to the external momentum are UV finite,
\begin{subequations}
\begin{align}\label{eq:dZUV}
\dZOL{11}\Big|_\text{div} &= - \Re\left[\Sigma_{11}^{(1)\prime}\right]_\text{div}, \\
\dZOL{22}\Big|_\text{div} &= - \Re\left[\Sigma_{22}^{(1)\prime}\right]_\text{div}, \\
\dZOL{12}\Big|_\text{div} &=  \dZOL{21}\Big|_\text{div} = 0,
\end{align}
\end{subequations}
where we used the subscript ``div'' to denote that only the UV divergent part is taken into account. We do not specify the momentum at which the derivatives of the self-energies are evaluated, since their UV divergent part does not depend on the external momentum.

As noted already above, the off-diagonal field renormalization constants are not needed to render Green's functions with external Higgs bosons UV finite. In the addition to the UV divergent part, we also allow for a UV finite contribution to the field renormalization constants. This part will be specified in \Sec{sec:04_poledet}. Note that the specification of the field renormalization constants also fixes the renormalization and therefore definition of $\tan\beta$.

In the approximation of vanishing electroweak gauge couplings and vanishing external momentum, as applied at the two-loop level, all two-loop field renormalization constants drop out of the calculation~\cite{Hollik:2014bua}. They also do not appear at higher-orders, since we work in the approximation of vanishing external momentum at the two-loop level. Therefore, we do not impose any condition on them in this work.


\subsubsection{Renormalized self-energies}

At the one-loop level, the renormalized self-energies are given in terms of the unrenormalized self-energies and counterterms by
\begin{subequations}
\begin{align}\label{eq:ren_seOL}
\hat\Sigma^{(1)}_{hh}(p^2) &= \Sigma^{(1)}_{hh}(p^2) - \dmOL{h}  + \frac{1}{2}\dZOL{hh}\left(p^2-m_h^2\right),\\
\hat\Sigma^{(1)}_{HH}(p^2) &= \Sigma^{(1)}_{HH}(p^2) - \dmOL{H}  + \frac{1}{2}\dZOL{HH}\left(p^2-m_H^2\right),\\
\hat\Sigma^{(1)}_{AA}(p^2) &= \Sigma^{(1)}_{AA}(p^2) - \dmOL{A}  + \frac{1}{2}\dZOL{AA}\left(p^2-M_A^2\right),\\
\hat\Sigma^{(1)}_{hH}(p^2) &= \Sigma^{(1)}_{hH}(p^2) - \dmOL{hH} + \frac{1}{2}\dZOL{hH}\left(p^2-\frac{1}{2}(m_h^2+m_H^2)\right),\\
\hat\Sigma^{(1)}_{hA}(p^2) &= \Sigma^{(1)}_{hA}(p^2) - \dmOL{hA} + \frac{1}{2}\dZOL{hA}\left(p^2-\frac{1}{2}(m_h^2+M_A^2)\right),\\
\hat\Sigma^{(1)}_{HA}(p^2) &= \Sigma^{(1)}_{HA}(p^2) - \dmOL{HA} + \frac{1}{2}\dZOL{HA}\left(p^2-\frac{1}{2}(m_H^2+M_A^2)\right), \\
\hat\Sigma^{(1)}_{\Hpm\Hpm}(p^2) &= \Sigma^{(1)}_{\Hpm\Hpm}(p^2) - \dmOL{\Hpm}  + \dZOL{\Hpm\Hpm}\left(p^2-m_{\Hpm}^2\right),
\end{align}
\end{subequations}
where the various $\dOL m^2$ denote the one-loop mass counterterms, which are given explicitly in \App{app:07_ren}. The renormalized self-energies involving Goldstone bosons are obtained analogously.

The renormalized two-loop self-energies are given by
\begin{align}\label{eq:renTLse}
\hat\Sigma^{(2)}_{ij}(0) &= \Sigma^{(2)}_{ij}(0) - \delta^{(2)}m^\mathbf{Z}_{ij},\hspace{.5cm}(i,j = h, H, A, G, \Hpm, \Gpm),
\end{align}
where $\delta^{(2)}m^\mathbf{Z}_{ij}$ is a combination of mass counterterms and field renormalization constants. Explicit expressions are listed in \App{app:07_ren}. Note that the unrenormalized two-loop self-energies contain subloop renormalization contributions.


\section{Momentum dependence of heavy contributions}
\label{sec:03_p2heavy}

In this Section, we review the decoupling behavior of heavy particles focusing especially on the momentum dependence induced by them. Our discussion is based on the proof of the decoupling theorem (see \cite{Appelquist:1974tg}) presented in \cite{Collins:1984xc}. The arguments in this Section are generally applicable and not restricted to the MSSM.

We investigate the kinematic behavior of a theory with one light boson with mass $m$ and one heavy boson with mass $M$ at energies $p^2$, assuming\footnote{We restrict us here to the case of one light and one heavy boson for simplicity. For the same reason, we also assume that all $n$-point vertex functions depend only on one external momentum. The arguments made in this Section, however, can easily be generalized to theories with a richer particle spectrum and more complicated kinematics.}
\begin{align}
m^2 \sim p^2 \ll M^2.
\end{align}
In order to study the momentum dependence of loop corrections involving the heavy particle, we consider the loop corrections to the unrenormalized $n$-point vertex function $\Gamma^{(n)}$ with external light fields only.

The mass dimension of $\Gamma^{(n)}$, depending on the number of external light bosons, is given by
\begin{align}
\left[\Gamma^{(n)}\right] = 4 - n,
\end{align}
where we assume that the mass dimension of the overall Lagrangian is four. The square brackets are used to denote the mass dimension of the quantity within the brackets.

Next, we separate the $n$-point vertex function into a light and a heavy part,
\begin{align}
\Gamma^{(n)} = \Gamma^{(n)}_\light(p^2, m) + \Gamma^{(n)}_\heavy(p^2, m, M),
\end{align}
where the light part is assumed to contain all loop corrections involving only the light particle; the heavy part, to contain loop corrections involving at least one heavy particle. The heavy part can be expressed as
\begin{align}
\Gamma^{(n)}_\heavy(p^2,m,M) = \sum_i \Gamma^{(n)}_{\heavy,i}(p^2,m,M) = \sum_i a_i^{(n)} I_i^{(n)}(p^2, m, M),
\end{align}
where the sum runs over all involved loop diagrams. The involved loop integrals are labelled by $I_i^{(n)}$. The $a_i^{(n)}$'s are dimensionless prefactors (e.g.\ gauge couplings).\footnote{For simplicity, we assume that the considered theory has no dimensionful couplings. A generalization to theories with dimensionful couplings is straightforward.} Consequently, the mass dimension of the integrals is the same as of the overall $n$-point vertex function,
\begin{align}
\left[I_i^{(n)}\right] = \left[\Gamma^{(n)}_\heavy\right] = 4 - n.
\end{align}
We want to know how this integral behaves in the limit of large $M$. Integrals can be expanded in the limit of an internal mass becoming heavy by using the technique of expansion by regions \cite{Tkachov:1991ev,Pivovarov:1991du,Smirnov:1990rz,Smirnov:1994tg,Smirnov:1996ng}. The leading contribution in this limit is given by the region in which all loop momenta are similar in size as the heavy mass $M$. In this region, all external momenta and light internal masses can be neglected. Consequently, the leading piece in the large mass limit can depend on neither the external momenta nor the light internal masses. By dimensional analysis, we obtain
\begin{align}
I_i^{(n)}(p^2, m^2, M^2) \xrightarrow{M \rightarrow \infty} b_{i,0}^{(n)} M^{[I_i^{(n)}]}  + \dots,
\end{align}
where the dots denote terms subleading in the $M\rightarrow\infty$ limit. $b_{i,0}^{(n)}$ is a dimensionless coefficient, which does neither depend on $m$ nor on $p^2$. It can, however, contain logarithms involving the heavy mass and the renormalization scale $\mu_R$.\footnote{We employ dimensional regularization for non-supersymmetric theories and dimensional reduction for supersymmetric theories.}

To investigate the momentum dependence, we take the derivative of the $n$-point vertex function with respect to the external momentum. The derivative lowers the mass dimension,
\begin{align}
\left[\left(\frac{d}{d p^2}\right)^k\Gamma^{(n)}_\heavy\right] = \left[\Gamma^{(n)}_\heavy\right] - 2 k, \hspace{.5cm} \left[\left(\frac{d}{d p^2}\right)^k I_i^{(n)}\right] = \left[I_i^{(n)}\right] - 2 k
\end{align}
with $k\ge 1$. Consequently, we have
\begin{align}
\left(\frac{d}{d p^2}\right)^k I_i^{(n)}(p^2, m^2, M^2) \xrightarrow{M \rightarrow \infty} b_{i,k}^{(n)} M^{[I_i^{(n)}] - 2 k},
\end{align}
and for the $n$-point vertex function
\begin{align}
\left(\frac{d}{d p^2}\right)^k \Gamma^{(n)}_\heavy \xrightarrow{M \rightarrow \infty}
   & \sum_i a_i^{(n)} b_{i,k}^{(n)} M^{[I_i^{(n)}] - 2 k} + \dots = \nonumber \\
={}& \sum_i a_i^{(n)} b_{i,k}^{(n)} M^{4 - n - 2 k} + \dots \; .
\end{align}
This implies that the leading contribution in the $M\rightarrow \infty$ limit is given by
\begin{align}
\left(\frac{d}{d p^2}\right)^k \Gamma_\heavy^{(n)} \xrightarrow{M \rightarrow \infty} 0,
\end{align}
if there are more than two external particles ($n>2$).

In case of only two external particles, there is a finite non-zero remainder if only one derivative is applied,
\begin{align}
\left(\frac{d}{d p^2}\right) \Gamma_\heavy^{(2)} \xrightarrow{M \rightarrow \infty} \sum_i a_i^{(n)} b_{i,1}^{(n)},
\end{align}
whereas
\begin{align}
\left(\frac{d}{d p^2}\right)^k \Gamma_\heavy^{(2)} \xrightarrow{M \rightarrow \infty} 0
\end{align}
for $k > 1$.

This means that in the limit $M\rightarrow \infty$, the heavy parts of $n$-point vertex functions do not depend on the external momentum if there are more than two external particles,
\begin{align}
\Gamma^{(n)}_\heavy(p^2,m,M)\big|_{M \rightarrow \infty} = \Gamma^{(n)}_\heavy(m,M)\big|_{M \rightarrow \infty}
\end{align}
for $n > 2$. If there are only two external particles, the dependence on the external momentum is given by
\begin{align}\label{eq:gamma2_exp}
\Gamma^{(2)}_\heavy(p^2,m,M)\big|_{M \rightarrow \infty} ={}& \Gamma^{(2)}_\heavy(p^2=m^2,m,M)\big|_{M \rightarrow \infty}  \nonumber\\
&+ (p^2 - m^2)\left(\frac{d}{d p^2}\Gamma^{(2)}_\heavy\right)(m,M)\big|_{M \rightarrow \infty},
\end{align}
where we expanded around $p^2 = m^2$. Note that we did not specify at which external momentum the term $\left(\frac{d}{d p^2}\Gamma^{(2)}_\heavy\right)(m,M)\big|_{M \rightarrow \infty}$ is evaluated, since it is independent of the external momentum.

The remaining dependence on the external momentum in the second line of \Eq{eq:gamma2_exp} is not affecting physical observables. In fact, it can be removed by renormalizing the two-point vertex function and absorbing the remaining dependence on the external momentum into the field renormalization of the light external boson (see \cite{Collins:1984xc} for more details),
\begin{align}\label{eq:dZHhOS}
\delta Z\big|_\text{fin} = - \left(\frac{d}{d p^2}\Gamma^{(2)}_\heavy\right)(m,M)\big|_{M \rightarrow \infty}.
\end{align}
The UV divergent part is chosen independently of the UV finite part such that the derivative of the two-point vertex function with respect to the external momentum is finite. In contrast to the OS scheme, in which the derivative of the full self-energy, including ``light'' and ``heavy'' contributions, is renormalized to zero, here only the ``heavy'' part is absorbed. Therefore, we will refer to this scheme as ``heavy-OS scheme'' in the following.

In general, field renormalization constants drop out in the calculation of physical observable if all corrections of a specific order are completely taken into account. Consequently, the momentum dependence of heavy contributions vanishes in the limit of the heavy mass going to infinity -- as expected from the decoupling theorem -- at each complete order independently of the chosen renormalization scheme, since the result has to be independent of this choice. At higher incomplete orders, we have to impose \Eq{eq:dZHhOS} to ensure that the momentum dependence of heavy contributions is canceled out.

Following a very similar reasoning, also the renormalization scale dependence of the two-point vertex function can be assessed. Analysing the degree of divergence after applying derivatives with respect to the external momentum shows: Leaving aside the intrinsic renormalization scale of the parameters entering the derivative of the two-point vertex function, also its explicit scale dependence can be absorbed into the Higgs field renormalization.


\section{Higgs pole determination in the MSSM}
\label{sec:04_poledet}

Next, we will discuss how the decoupling behavior of heavy particles, as discussed in \Sec{sec:03_p2heavy}, affects the determination of the Higgs-boson propagator poles in the MSSM. Note that the discussion, despite being specific to the MSSM here, is actually applicable to general models involving large mass hierarchies between two or more particles.

The Higgs-boson propagator poles, whose real parts are the squared physical masses, are obtained by solving the equations
\begin{subequations}\label{eq:poleeqs}
\begin{align}
\det\left(\mathbf{\hat\Gamma}_{hHA}(p^2)\right) &= 0, \\
\det\left(\hat\Gamma_{\Hpm\Hpm}(p^2)\right)     &= 0,
\end{align}
\end{subequations}
with the two-point vertex functions
\begin{subequations}
\begin{align}\label{eq:propmatrix}
\mathbf{\hat\Gamma}_{hHA}(p^2) &= i
\begin{pmatrix}
p^2 - m_h^2 + \hat\Sigma_{hh}(p^2) & \hat\Sigma_{hH}(p^2)               & \hat\Sigma_{hA}(p^2)              \\
\hat\Sigma_{hH}(p^2)               & p^2 - m_H^2 + \hat\Sigma_{HH}(p^2) & \hat\Sigma_{HA}(p^2)              \\
\hat\Sigma_{hA}(p^2)               & \hat\Sigma_{HA}(p^2)               & p^2 - M_A^2 + \hat\Sigma_{AA}(p^2)
\end{pmatrix}, \\
\hat\Gamma_{\Hpm\Hpm}(p^2) &= i\left( p^2 - m_{\Hpm}^2 + \hat\Sigma_{\Hpm\Hpm}(p^2) \right).
\end{align}
\end{subequations}
We neglect mixing with the Goldstone and the $Z$ boson, since these induce only subleading two-loop contributions. In order to keep the expressions short, we do not explicitly write down the contribution coming from the EFT calculation but assume that it is intrinsically contained in the self-energies.

In the following, we will discuss different methods for solving \EqsSub{eq:poleeqs}.


\subsection{Numerical pole determination with \DR field renormalization}
\label{subsec:num_poledet}

The most direct approach to determine the Higgs-boson propagator poles is to solve \EqsSub{eq:poleeqs} numerically. This can be done employing different schemes for the Higgs field renormalization. First, we discuss the case of \DR Higgs field renormalization, which was used as default scheme in \FH until version \texttt{2.14.2}. This means that no UV finite parts are added to the UV divergent parts as specified in \EqsSub{eq:dZUV}.

In~\cite{Bahl:2017aev}, this procedure was closely investigated in the limit of $M_A \gg M_Z$. In this limit, \mbox{$h$-$H$}~mixing effects in the \cp-even Higgs sector are suppressed by powers of $M_A$ and the mass of the $h$ boson can be obtained by solving the simpler equation
\begin{align}
p^2 - m_h^2 + \hat\Sigma_{hh}(p^2)= 0
\end{align}
up to corrections suppressed by $M_A$. Solving this equation up to the two-loop order yields
\begin{align}\label{Mh2_TL_DecouplingLimit_Eq}
M_h^2 ={}& m_h^2 - \hat\Sigma_{hh}^{(1)}(m_h^2) - \hat\Sigma_{hh}^{(2)}(0)\gl + \hat\Sigma_{hh}^{(1)\prime}(m_h^2) \hat\Sigma_{hh}^{(1)}(m_h^2) = \nonumber\\
={}& m_h^2 - \hat\Sigma_{hh}^{(1)}(m_h^2) - \hat\Sigma_{hh}^{(2)}(0)\gl \nonumber\\
& + \left(\hat\Sigma_{hh}^{\heavy,(1)\prime}(m_h^2)+\hat\Sigma_{hh}^{\light,(1)\prime}(m_h^2)\right) \hat\Sigma_{hh}^{(1)}(m_h^2),
\end{align}
where the number in the superscript is used to indicate the loop order. Again, the subscript ``gl'' is used to indicate the approximation of vanishing electroweak gauge couplings at the two-loop level. In the last step, we split up the derivative of the $hh$ self-energy into a part containing all contributions involving heavy particles (omitting terms suppressed by the heavy mass scale), labelled by the superscript ``heavy'' particles, and a part containing all contributions from light particles (including terms suppressed by a heavy mass scale), labelled by the superscript ``light''. Note that, in order to simplify the notation in this Section, the labels have a slightly different meaning in this Section as compared to \Sec{sec:03_p2heavy}. Here, the light contribution also includes terms suppressed by a large mass, whereas these terms are omitted in the heavy contribution. In~\cite{Bahl:2017aev}, a scenario with all non-SM particles sharing one common mass scale was assumed. Therefore, $\hat\Sigma_{hh}^{\light}$ contains contributions from SM particles as well as terms suppressed by a heavy non-SM mass.

The analysis in~\cite{Bahl:2017aev} showed that the term labelled by ``heavy'' in \Eq{Mh2_TL_DecouplingLimit_Eq} is cancelled by parts of the subloop renormalization contained in the two-loop self-energy. The discussion in \Sec{sec:03_p2heavy} clearly shows the reason for this cancellation: The momentum dependence of heavy contributions to two-point vertex functions, i.e.\ the term proportional to $\hat\Sigma_{hh}^{\heavy,(1)\prime}(m_h^2)$ in \Eq{Mh2_TL_DecouplingLimit_Eq}, could be absorbed into a field renormalization constant and therefore drops out order by order in the perturbative expansion.\footnote{Note that we included terms suppressed by a heavy mass, which can contain additional momentum-dependent terms, into the ``light'' contribution in this Section.} If the Higgs field renormalization is fixed in the \DR scheme, it is therefore important for a proper cancellation to ensure that all corrections at the two-loop level are included at the same level of accuracy and that no higher order terms proportional to $\hat\Sigma_{hh}^{\heavy,(1)\prime}(m_h^2)$ are induced. Determining the Higgs-boson propagator poles numerically, however, does not allow to easily control the inclusion or exclusion of certain terms.

Concerning the corrections included in \FH, this means that in order to achieve a complete cancellation at the two-loop level only \order{\alt^2,\alt\alb,\alb^2} corrections should be taken into account for the term $\hat\Sigma_{hh}^{\heavy,(1)\prime}(m_h^2) \hat\Sigma_{hh}^{(1)}(m_h^2)$, since the two-loop self-energy is included up to this order.


\subsection{Fixed-order pole determination}
\label{subsec:fo_poledet}

As an approach to solve this problem, it was suggested in \cite{Bahl:2017aev} to simply employ \Eq{Mh2_TL_DecouplingLimit_Eq} and to ensure that the last term in the equation is computed at the same level of accuracy as the two-loop self-energy. This simple procedure is, however, only applicable in case of high $M_A$ for which \mbox{$h$-$H$}~mixing effects are negligible. Here, we work out a method based upon the same idea -- solving \EqsSub{eq:poleeqs} at a fixed-order -- applicable also for low $M_A$. This method was used in the \FH versions~\texttt{2.14.0} through~\texttt{2.14.2}.

A proper cancellation is guaranteed by including all two-loop corrections at the same level of accuracy. In case of low $M_A$, we cannot neglect the off-diagonal elements of the Higgs two-point vertex function, since their inclusion is crucial for a precise prediction. Instead, we consider each element of the two-point vertex function separately.

The complex pole corresponding to the $i$-th boson ($\mbox{i = 1,2,3}$) is given at the one-loop level by
\begin{align}\label{eq:oneloopsol}
\left(\mathcal{M}_{i}^{(1)}\right)^2 = m_{i}^2 - \hat\Sigma_{ii}^{(1)}(m_{i}^2),
\end{align}
where the superscript indicates the loop-order. We choose the tree-level masses to be
\begin{align}
m_{1}^2 &= m_h^2,\quad m_{2}^2 = m_H^2,\quad m_{3}^2 = M_A^2
\end{align}
and the one-loop self-energies to be
\begin{align}
\hat\Sigma_{{1}{1}}^{(1)} &= \hat\Sigma_{hh}^{(1)},\quad \hat\Sigma_{{2}{2}}^{(1)} = \hat\Sigma_{HH}^{(1)},\quad \hat\Sigma_{{3}{3}}^{(1)} = \hat\Sigma_{AA}^{(1)}.
\end{align}
Expanding each of the self-energies appearing in the two-point vertex function matrix (see \Eq{eq:propmatrix}) around the one-loop solution of \Eq{eq:oneloopsol} up to the two-loop level yields the expanded two-point vertex function matrix $\hat\Gamma_{hHA}^{i-\text{exp}}$ with the matrix elements given by
\begin{align}
 \left(\hat\Gamma_{hHA}^{i-\text{exp}}(p^2)\right)_{jk} &= D_{jk}(p^2) + \hat\Sigma_{jk}^{(1)}(m_{i}^2) + \hat\Sigma_{jk}^{(2)}(0)\gl  - \left[\hat\Sigma_{jk}^{(1)\prime}(m_{i}^2)\hat\Sigma_{{i}{i}}^{(1)}(m_{i}^2)\right]_\text{gl}
\end{align}
with $j,k = 1,2,3$ and
\begin{align}
D_{jk}(p^2) = (p^2 - m_j^2)\delta_{jk}.
\end{align}
The subscript of the reducible terms $\hat\Sigma^{(1)\prime}\hat\Sigma^{(1)}$, ``gl'', indicates that we calculate these in the gaugeless limit (i.e.\ vanishing electroweak gauge couplings corresponding to the approximation used at the two-loop level). Consequently, these terms are of \order{\alt^2,\alt\alb,\alb^2} and therefore of the same order as the corresponding parts of the two-loop self-energies $\hat\Sigma^{(2)}$. Thereby, it is guaranteed that cancellations between the various two-loop contributions are correctly taken into account.

The next step is to find the zeroes of $\hat\Gamma_{hHA}^{i-\text{exp}}(p^2)$. We pick the pole with the largest overlap of the associated physical state with the tree-level mass eigenstate. This pole corresponds to the one-loop solution given in \Eq{eq:oneloopsol}. The physical mass squared is then given by taking the real part of this pole. The mass of the $\Hpm$ boson is determined accordingly. This procedure successfully avoids inducing higher order terms which would cancel in a more complete calculation.

However, as we will argue in \Sec{subsec:poledet_comp} and explicitly show in \Sec{sec:05_results}, the approach presented in this Section yields unsatisfactory results close to crossings points where two mixing Higgs bosons are almost mass degenerate. This issue does not appear if the propagator poles are determined numerically.


\subsection{Numerical pole determination with heavy-OS field renormalization}
\label{subsec:hOS_poledet}

Last, we present a third approach to determine the Higgs-boson propagator poles avoiding the problems of the approaches presented above. As in \Sec{subsec:num_poledet}, the Higgs-boson propagator poles are determined numerically. However, to avoid the problem with uncancelled terms beyond the order of the fixed-order calculation, we do not employ the \DR scheme for the Higgs field renormalization.

As shown in \Sec{sec:03_p2heavy}, the Higgs field renormalization can be used to absorb the momentum dependence of contributions from heavy particles. As discussed in \Sec{subsec:num_poledet}, their momentum dependence is the origin of the uncancelled higher-order terms. Consequently, the issue is resolved by fixing the Higgs field renormalization according to the heavy-OS scheme (as presented in \Sec{sec:03_p2heavy}). This means that the one-loop Higgs field renormalization constants (introduced in \Sec{sec:02_MSSMintro}) are fixed by
\begin{align}\label{eq:MSSMhOS}
\delta^{(1)}Z_{ij}\Big|_\text{hOS} &= \delta^{(1)}Z_{ij}\Big|_\DR - \Sigma_{ij}^{\heavy,(1)\prime}\Big|_\text{fin} - \gamma_{ij}^{\light,(1)}\ln\frac{\mu_R^2}{Q_\text{pd}^2}\hspace{.3cm}(i,j=h,H,A).
\end{align}
Note that we did not specify at which external momentum the second term is evaluated, since it does not depend on the external momentum. In addition to absorbing the momentum dependence of the heavy contributions, we also add a term proportional to the Higgs anomalous dimension involving light fields only,\footnote{Terms suppressed by a heavy scale do not contribute to the Higgs anomalous dimensions.}
\begin{align}\label{eq:anomal_dim_light}
\gamma_{ij}^{\light,(1)} = \frac{\partial}{\partial \ln \mu_R^2} \Sigma_{ij}^{\light,(1)\prime}\Big|_{\mu_R = Q_\text{pd}},
\end{align}
which is used to absorb the dependence of Higgs self-energies' derivatives on the renormalization scale $\mu_R$ (see end of \Sec{sec:03_p2heavy}). In this way, we separate the scale dependence associated with the pole determination from the renormalization scale dependence of the fixed-order corrections. The corresponding reference scale at which the poles are determined is called~$Q_\text{pd}$. In order to avoid inducing large logarithms via the pole determination we choose it equal to the OS top mass, $Q_\text{pd}=M_t$.

Explicit expressions for the one-loop renormalization constants in the gauge eigenstate basis are listed in \App{app:09_finfieldren}.

Since the fixed-order calculation used by default in \FH applies the approximation of vanishing external momentum at the two-loop level, no finite pieces have to be added to the two-loop field renormalization constants. Going beyond this approximation (see~\cite{Borowka:2014wla,Borowka:2015ura,Borowka:2018anu}) implies that also the two-loop field renormalization constants should be fixed in the heavy-OS scheme.

Having calculated the Higgs self-energies using the field renormalization scheme specified in \Eq{eq:MSSMhOS}, the Higgs-boson propagator poles are then determined by solving \EqsSub{eq:poleeqs} numerically.

Note that employing the full OS scheme, meaning that the full derivatives of the Higgs self-energies are renormalized to be zero, is not a good alternative. In the OS scheme, unphysical threshold effects coming from the ``light'' contributions are induced (see e.g.~\cite{Frank:2002qf}).

It is important to note that the renormalization scheme choice made in \Eq{eq:MSSMhOS} affects the definition of $\tan\beta$ (see \Sec{sec:02_MSSMintro}). In particular $\tan\beta$ is not anymore a MSSM quantity but rather a Two-Higgs-Doublet Model (THDM) one. This is explained in detail in \cite{Bahl:2018jom}. By choosing the field renormalization as in \Eq{eq:MSSMhOS}, $\tan\beta$ is defined at the scale $Q_\text{pd}$, which is set to the OS top mass. In situations where the THDM scale $M_A$ is much larger than the top-quark mass, this definition is not well-suited. Furthermore, it would be preferable to have an MSSM $\tan\beta$ as input for a MSSM calculation. We present two methods to address these shortcomings in \App{app:08_tbrepara}.

In addition to the Higgs poles, also the $\mathbf{Z}$-matrix relating the tree-level mass eigenstates and the external physical states in scattering amplitudes is calculated~\cite{Chankowski:1992er,Heinemeyer:2001iy,Frank:2006yh,
Fuchs:2016swt,Fuchs:2017wkq,Domingo:2017rhb}. This matrix depends on the choice of the Higgs field renormalization. For other calculations (e.g.~decay rates or productions cross-sections), the heavy-OS scheme is, however, not needed to avoid uncancelled higher order terms, since these processes involve more than two external legs. In order to ease the use of the output $\mathbf{Z}$-matrix, we therefore discuss how it can be transferred to the \DR scheme, which is more widely used, in \App{app:10_Zmatrix}.


\subsection{Comparison of the different methods}
\label{subsec:poledet_comp}

To further illustrate the difference between the various approaches, we investigate the corrections to the mass of the $h$ boson up to the two-loop level.

In the first approach, numerical pole determination with \DR Higgs field renormalization, they are given by
\begin{align}\label{eq:Mh_pd1}
\left(M_h^2\right)_\text{num,\DR} ={}& m_h^2 - \hat\Sigma_{hh}^{(1)}(m_h^2)\big|_{\delta Z|_\DR} - \hat\Sigma_{hh}^{(2)}(0)\big|_{\text{gl},\delta Z|_\DR} + \hat\Sigma_{hh}^{(1)\prime}(m_h^2)\big|_{\delta Z|_\DR} \hat\Sigma_{hh}^{(1)}(m_h^2)\big|_{\delta Z|_\DR} \nonumber\\
&+ \frac{\left(\hat\Sigma_{hH}^{(1)}(m_h^2)\big|_{\delta Z|_\DR}\right)^2}{m_h^2 - m_H^2} + \frac{\left(\hat\Sigma_{hA}^{(1)}(m_h^2)\big|_{\delta Z|_\DR}\right)^2}{m_h^2 - M_A^2},
\end{align}
where we added the subscript ``$\delta Z|_\DR$'' to clarify that the field renormalization constants are fixed in the \DR scheme.

In the fixed-order approach, we obtain
\begin{align}\label{eq:Mh_pd2}
\left(M_h^2\right)_\text{FO} ={}& m_h^2 - \hat\Sigma_{hh}^{(1)}(m_h^2)\big|_{\delta Z|_\DR} - \hat\Sigma_{hh}^{(2)}(0)\big|_{\text{gl},\delta Z|_\DR} + \hat\Sigma_{hh}^{(1)\prime}(m_h^2)\big|_{\text{gl},\delta Z|_\DR} \hat\Sigma_{hh}^{(1)}(m_h^2)\big|_{\text{gl},\delta Z|_\DR} \nonumber\\
&+ \frac{\left(\hat\Sigma_{hH}^{(1)}(m_h^2)\big|_{\delta Z|_\DR}\right)^2}{m_h^2 - m_H^2} + \frac{\left(\hat\Sigma_{hA}^{(1)}(m_h^2)\big|_{\delta Z|_\DR}\right)^2}{m_h^2 - M_A^2}.
\end{align}
We see that the fixed-order pole determination avoids including the terms identified in \Sec{subsec:num_poledet}, which would cancel in a more complete calculation.

Using numerical pole determination with OS pole determination, we get
\begin{align}\label{eq:Mh_pd3}
\left(M_h^2\right)_\text{num,hOS} ={}& m_h^2 - \hat\Sigma_{hh}^{(1)}(m_h^2)\big|_{\delta Z_\text{hOS}} - \hat\Sigma_{hh}^{(2)}(0)\big|_{\text{gl},\delta Z_\text{hOS}} + \hat\Sigma_{hh}^{(1)\prime}(m_h^2)\big|_{\delta Z_\text{hOS}} \hat\Sigma_{hh}^{(1)}(m_h^2)\big|_{\delta Z_\text{hOS}} \nonumber\\
&+ \frac{\left(\hat\Sigma_{hH}^{(1)}(m_h^2)\big|_{\delta Z_\text{hOS}}\right)^2}{m_h^2 - m_H^2} + \frac{\left(\hat\Sigma_{hA}^{(1)}(m_h^2)\big|_{\delta Z_\text{hOS}}\right)^2}{m_h^2 - M_A^2},
\end{align}
where we added the subscript ``$\delta Z_\text{hOS}$'' to clarify that the field renormalization constants are fixed in the heavy-OS scheme. Again, terms which would cancel in a more complete calculation are not included.

The difference between the numerical pole determination employing heavy-OS Higgs field renormalization and the fixed-order pole determination is given by\footnote{Note that $\hat\Sigma_{hh}^{(1)}(m_h^2)$ does not depend explicitly on the scheme chosen for the Higgs field renormalization. Assuming the same definition of $\tan\beta$, $\hat\Sigma_{hh}^{(1)}(m_h^2)$ is therefore equal for \DR Higgs field renormalization and for heavy-OS Higgs field renormalization.}
\begin{align}\label{eq:diffpd12}
\left(M_h^2\right)_\text{num,hOS} - \left(M_h^2\right)_\text{FO} ={}& \Sigma_{hh}^{\light,(1)\prime}(m_h^2)\big|_\text{fin}\cdot\hat\Sigma_{hh}^{(1)}(m_h^2) \nonumber\\
&- \left[\Sigma_{hh}^{\light,(1)\prime}(m_h^2)\big|_\text{fin}\cdot\hat\Sigma_{hh}^{(1)}(m_h^2)\right]_\text{gl} .
\end{align}
We see that the numerical pole determination employing heavy-OS renormalization includes additional terms. The term containing the light contribution to the derivative of the $hh$~self-energy and the $hh$~self-energy is included fully and not only in the limit of vanishing electroweak gauge couplings. The electroweak next-to-leading logarithms which are contained in this term are of the same order as some of the logarithms resummed by the means of the EFT calculation, which are added on top of the fixed-order result. Therefore, this term, which was missed in \cite{Bahl:2017aev}, should be included fully and not only in the limit of vanishing electroweak gauge couplings.

\medskip

In addition to this differences at the two-loop level, there are also differences at higher orders. In the case of fixed-order pole determination, unphysical discontinuities in the prediction of the Higgs boson masses close to crossings points, where two mixing Higgs bosons are almost mass degenerate, appear (numerical results displaying this behavior will be shown in \Sec{sec:05_results}). At these crossing points, a specific property of the Higgs-boson propagator matrix becomes important: Every complex pole is not only the solution of \EqsSub{eq:poleeqs} but also a pole of every element of the Higgs-boson propagator matrix~\cite{Fuchs:2016swt}. This property is fulfilled in case of numerical pole determination. In case of the fixed-order pole determination, this property, however, holds only up to the level of the truncation, i.e., the two-loop level. The violation at higher orders is the origin of the discontinuities in the Higgs mass predictions.


\section{Numerical results}
\label{sec:05_results}

In this Section, we will numerically compare the different methods to determine the Higgs-boson propagator poles. If not stated otherwise, $\tan\beta$ is defined in the \DR scheme at the scale $M_t$. If the Higgs field renormalization is fixed in the heavy-OS scheme, this is achieved by procedure outlined in \App{app:08_tbrepara}. To ease the comparison between the different results, we always employ the SM as EFT below the SUSY scale.

\begin{figure}\centering
\begin{minipage}{.48\textwidth}\centering
\includegraphics[width=\textwidth]{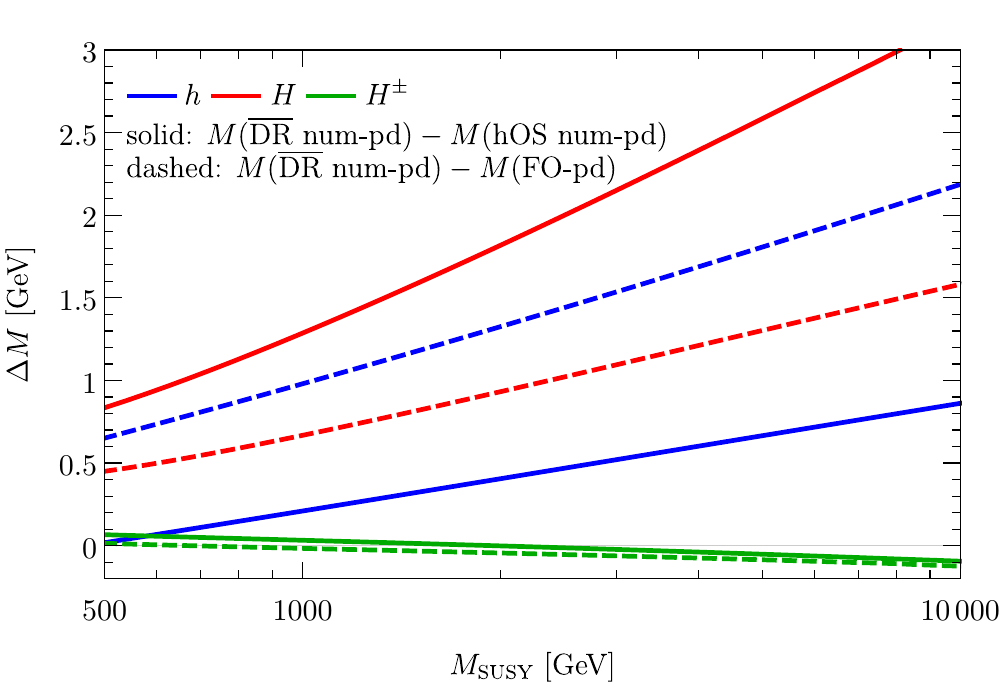}
\end{minipage}
\begin{minipage}{.48\textwidth}\centering
\includegraphics[width=\textwidth]{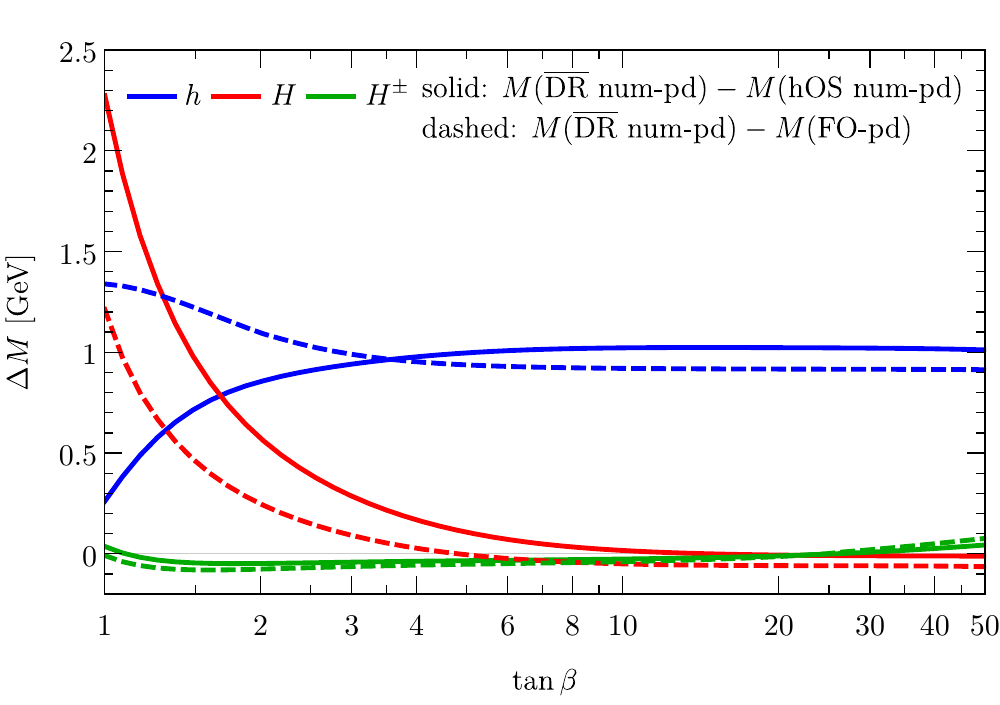}
\end{minipage}
\caption{Difference between the prediction of the Higgs boson masses using numerical pole determination with \DR Higgs field renormalization and heavy-OS Higgs field renormalization (solid). Also the difference between the results using numerical pole determination with \DR Higgs field renormalization and fixed-order pole determination is shown (dashed). Left: Variation of $\msusy$ setting $\tan\beta=1.1$. Right: Variation of $\tan\beta$ setting $\msusy = 2 \tev$.}
\label{fig:diff}
\end{figure}

First, we investigate scenarios with $M_A = 200 \gev$. All other soft-SUSY breaking masses are chosen to be equal to the common mass scale $\msusy$, which we vary between $0.5\tev$ and $10\tev$. The trilinear soft-breaking couplings are set equal to zero apart from $A_t$, which is fixed by setting $X_t^\OS/\msusy = 2$.\footnote{Here, we employ the OS scheme for the renormalization of the stop sector.} All parameters are chosen to be real.

Note that large parts of this parameter space are experimentally excluded due to the small $M_A$ value (see e.g.\ \cite{Bahl:2018zmf}). Here, we, however, only want to explore the numerical size of the differences between the various pole determination methods and do not perform a phenomenological analysis.

In the left plot of \Fig{fig:diff}, we vary $\msusy$ setting $\tan\beta = 1.1$ (this choice maximizes the difference between the pole determination methods). Shown are the differences between numerical pole determination with \DR Higgs field renormalization and heavy-OS field renormalization (solid) as well as between numerical pole determination with \DR Higgs field renormalization and fixed-order pole determination (dashed). The differences in the prediction of $M_h$ are shown in blue, of $M_H$ in red and of $M_\Hpm$ in green.

It is clearly visible that the shifts between the different pole determination methods rise logarithmically with increasing $\msusy$. Employing heavy-OS Higgs field renormalization leads to a downshift of $M_h$ of up to $\sim 1\gev$ with respect to the numerical pole determination employing \DR Higgs field renormalization and to a downwards shift of $M_H$ of up to \mbox{$\sim 3.5\gev$}. If instead the fixed-order pole determination is used, the shift of $M_h$ is enlarged by up to \mbox{$\sim 1\gev$}, whereas the shift of $M_H$ shrinks by up to $\sim 2\gev$. The different slopes obtained by using the different methods can be explained by next-to-leading logarithms missed in the case of fixed-order pole determination (see discussion in \Sec{subsec:poledet_comp}). The remaining differences between the two methods are caused by the different treatment of Higgs mixing effects, which are especially relevant for low $\tan\beta$. The impact of the different pole determination methods on $M_\Hpm$ is almost completely negligible.

In the right panel of \Fig{fig:diff}, we show the dependence of the shifts between the various pole determination methods on $\tan\beta$ for $\msusy = 2\tev$. The lines correspond to those of the left plot of \Fig{fig:diff}. For $\tan\beta \gtrsim 3$, the shifts of $M_h$ are approximately constant and equal to each other. In the region $\tan\beta\lesssim 3$, the difference between the results for $M_h$ using numerical pole determination with heavy-OS Higgs field renormalization and fixed-order pole determination rises when lowering $\tan\beta$. This behavior reflects the decreasing importance of Higgs mixing effects if $\tan\beta$ is increased. Due to the suppression of the top Yukawa coupling of the $H$~boson for high $\tan\beta$, the shifts of $M_H$ quickly decrease if $\tan\beta$ is raised. The shifts of $M_\Hpm$ show no significant dependence on $\tan\beta$.

\medskip

\begin{figure}\centering
\begin{minipage}{.48\textwidth}\centering
\includegraphics[width=\textwidth]{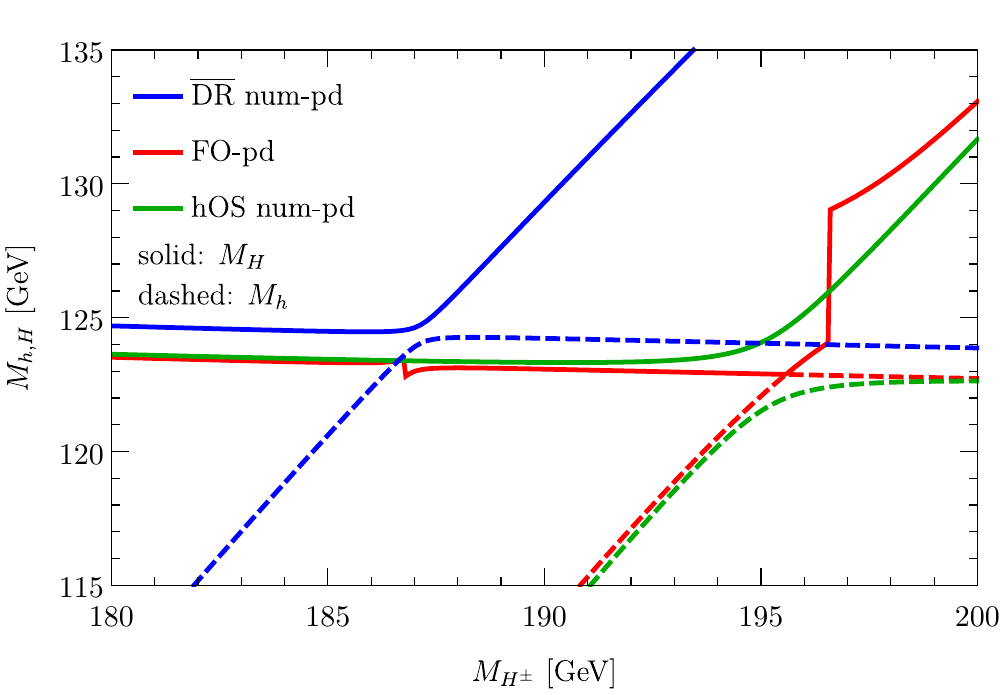}
\end{minipage}
\begin{minipage}{.48\textwidth}\centering
\includegraphics[width=\textwidth]{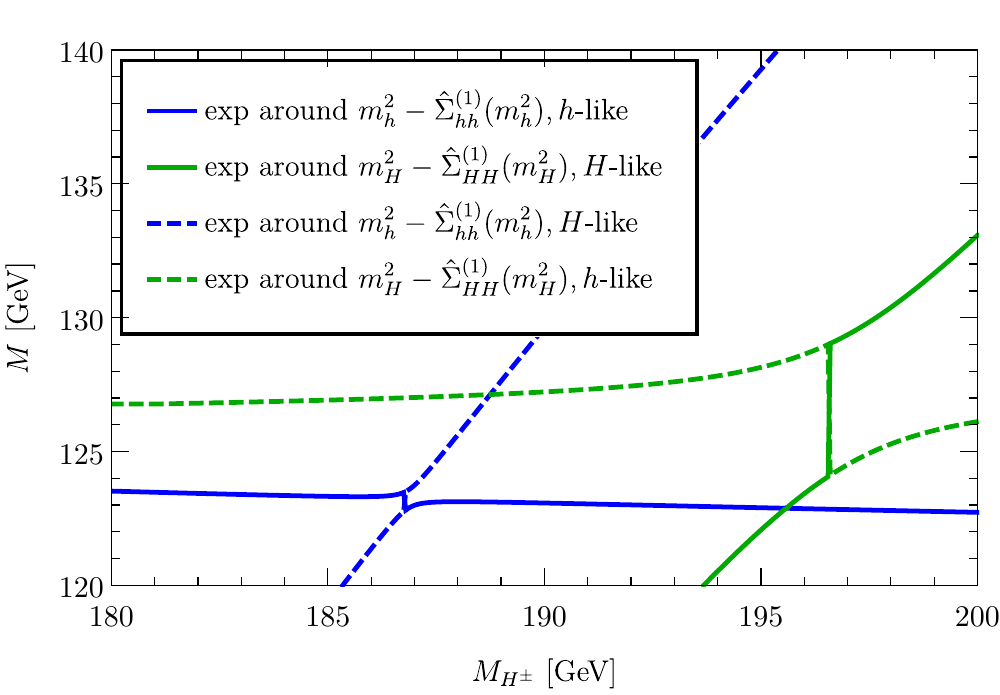}
\end{minipage}
\caption{Investigation of the $M_H^{125}$ benchmark scenario \cite{Bahl:2018zmf} setting $\tan\beta = 5.5$. The results for the mass of the lighter \cp-even Higgs boson are presented using dashed lines; the results for the mass of the heavier \cp-even Higgs boson using solid lines. Left: Comparison of the different pole determination methods. Right: Poles obtained using the fixed-order pole determination method.}
\label{fig:MH125}
\end{figure}

Next, we investigate the $M_H^{125}$ benchmark scenario proposed in~\cite{Bahl:2018zmf}. It is defined such that the second lightest \cp-even Higgs boson is SM-like and plays the role of the Higgs boson discovered at the LHC. Whereas we refer to~\cite{Bahl:2018zmf} for the exact definition of the scenario, we want to point out the large $\mu/\msusy$ value of $\sim 8$ with $\msusy\sim 750\gev$. We set $\tan\beta = 5.5$ and only vary $M_\Hpm$, which is used as input mass, as a free parameter.

The left plot of \Fig{fig:MH125} shows the results in these scenarios using the three methods to determine the Higgs-boson propagator poles presented in \Sec{sec:04_poledet} (results for $M_h$ are shown using dashed lines and for $M_H$ using solid lines). The numerical pole determination with \DR field renormalization (blue) has a crossing point, where $M_h$ and $M_H$ are almost equal, at $M_\Hpm \sim 187\gev$. This crossing point is shifted to $M_\Hpm \sim 195 \gev$ if the Higgs field renormalization constants are fixed in the heavy-OS scheme (green). This results in large shifts of the the two Higgs masses of up to $\sim 10 \gev$ in the region $187\gev \le M_\Hpm \le 195\gev$. These shifts are caused by terms which are induced by the pole determination using a \DR Higgs field renormalization and which would cancel in a more complete calculation. There size is especially large here because of the large $\mu/\msusy$ value. In most parts of the MSSM parameter space, where the lightest \cp-even Higgs is SM-like, the shifts between different pole determination methods are much smaller.

In the left plot of \Fig{fig:MH125}, also results using the fixed-order pole determination method are shown (red). The curve for $M_H$ has discontinuities at $M_\Hpm \sim 186.5 \gev$ and ${M_\Hpm \sim 196 \gev}$ by $\sim 1\gev$ and $\sim 5\gev$, respectively. We investigate these discontinuities more closely in the right plot of \Fig{fig:MH125}. As explained in \Sec{subsec:fo_poledet}, in the fixed-order pole determination approach each element of the inverse Higgs-boson propagator matrix is expanded around the one-loop solutions for the poles. In the case of real input parameters, these are $m_h^2 - \hat\Sigma_{hh}^{(1)}(m_h^2)$ (blue curves) and $m_H^2 - \hat\Sigma_{HH}^{(1)}(m_H^2)$ (green curves). After determining the eigenvalues of these expanded matrices, the overlap of the corresponding eigenstates with the tree-level mass eigenstates are determined.\footnote{Note that in this scenario, the conventional labelling of the masses, $M_h$ for the lighter mass and $M_H$ as heavier mass eigenstate can be confusing, since in some part of the parameter space the mass eigenstate corresponding to $M_h$ is mostly $H$-like and vice versa.}  For the solid curves, the eigenstate overlaps mostly with the tree-level mass eigenstate corresponding to the one-loop solutions around which we expanded. These are the correctly chosen eigenvalues and correspond to the red curve in the left plot of \Fig{fig:MH125}. The dashed curves show the wrongly chosen eigenvalues whose eigenstate overlaps mostly with the tree-level mass eigenstate not corresponding to the one-loop solutions. We observe that the discontinuities visible in the left plot of \Fig{fig:MH125} do not originate from numerical instabilities but are a consequence of the expansion of the propagator matrix at the two-loop level, which generates two different solution branches for the pole determination.

For $M_\Hpm \lesssim 186 \gev$ and $M_\Hpm \gtrsim 197 \gev$, the results using fixed-order pole determination and numerical pole determination with heavy-OS Higgs field renormalization are in good agreement. This confirms that both methods avoid inducing higher order terms which would cancel in a more complete calculation and which are included in the numerical pole determination employing \DR Higgs field renormalization. In the crossing point region, the numerical pole determination with heavy-OS Higgs field renormalization yields no unphysical discontinuities and is therefore superior to the fixed-order pole determination method. Consequently, this method was used for the definition of not only the $M_H^{125}$ scenario but also of the other benchmark scenarios presented in \cite{Bahl:2018zmf}.\footnote{One of the scenarios in \cite{Bahl:2018zmf} includes \cp-violation and serves as an example of the application of the pole determination with heavy-OS Higgs field renormalization to the case of $3\times 3$ mixing.}

\medskip

\begin{figure}\centering
\begin{minipage}{.48\textwidth}\centering
\includegraphics[width=\textwidth]{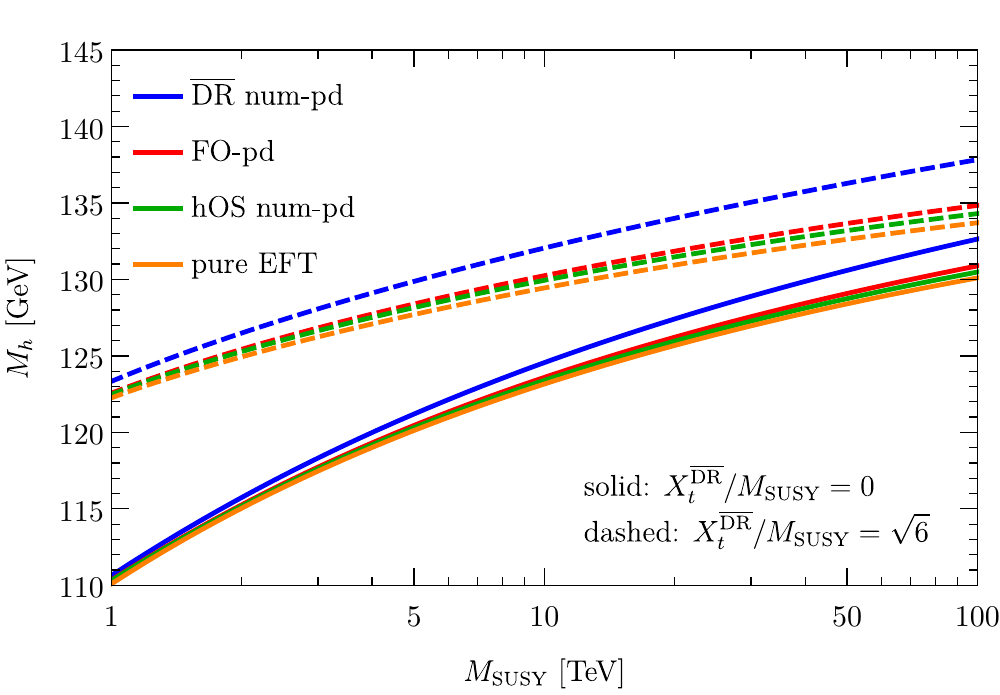}
\end{minipage}
\begin{minipage}{.48\textwidth}\centering
\includegraphics[width=\textwidth]{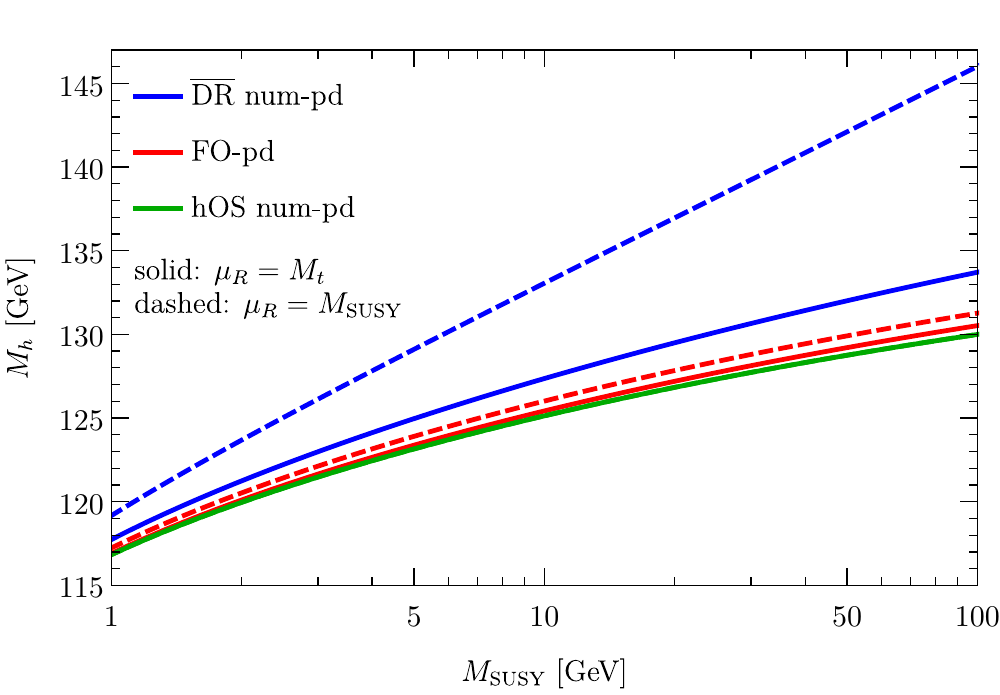}
\end{minipage}
\caption{Left: Comparison between the predictions for $M_h$ employing the different pole determination methods. In addition, the prediction of the pure EFT calculation is shown. Right: Comparison between the different pole determination methods for different choices of the renormalization scale $\mu_R$. $M_h$ is shown as function of $\msusy$ setting ${X_t^\OS/\msusy = 2}$ and $\tan\beta=4$.}
\label{fig:HS_scalevar}
\end{figure}

After investigating scenarios with light non-SM Higgs boson masses, we turn to a high-scale scenario where all non-SM masses are set equal to $\msusy$ (i.e., $M_A = \msusy$), which is varied between 1~TeV and 100~TeV. Here, we fix the parameters of the stop sector in the \DR scheme.

In the left plot of \Fig{fig:HS_scalevar}, we show the prediction for $M_h$ obtained using numerical pole determination with \DR Higgs field renormalization (blue), fixed-order pole determination (red) and numerical pole determination with heavy-OS Higgs field renormalization (green). In addition, we display the pure EFT result obtained with \FH (orange). All curves are shown for $X_t^\DR/\msusy=0$ (solid) and $X_t^\DR/\msusy=\sqrt{6}$ (dashed).

In the considered range for $\msusy$, terms suppressed by the SUSY scale are negligible and we expect to see a good agreement between the EFT prediction and the hybrid approach implemented in \FH. We observe that the remaining numerical difference between the EFT result and the hybrid approach is lowest if numerical pole determination with heavy-OS field renormalization is employed. In comparison to the fixed-order pole determination, especially the terms isolated in \Eq{eq:diffpd12} improve the agreement between the EFT and the hybrid result. The still remaining difference ($\sim 0.2\gev$ for vanishing stop mixing and $\sim 0.5\gev$ for $X_t^\DR/\msusy=\sqrt{6}$) can be explained by missing contributions proportional to the bottom Yukawa coupling in the EFT calculation and a different parameterization of non-logarithmic terms \cite{Bahl:2017aev}.

\medskip

Finally, we investigate the dependence of the results on the renormalization scale $\mu_R$. As mentioned above, for the fixed-order calculation of \FH the OS scheme is employed by default. Only the Higgsino mass parameter $\mu$ and, if the Higgs field renormalization is not fixed in the heavy-OS scheme, $\tan\beta$ are renormalized using the \DR~scheme. Consequently, the results for $M_h$ should only depend on the renormalization scale via the dependence of $\mu$ and possibly $\tan\beta$.

As argued in \Sec{sec:04_poledet}, the numerical pole determination method using \DR Higgs field renormalization leads to the inclusion of higher order terms depending on the renormalization scale $\mu_R$ which would cancel in case of a more complete fixed-order calculation. This incomplete cancellation is reflected by an enhanced dependence on the renormalization scale.

We illustrate this behavior in right plot of \Fig{fig:HS_scalevar}. In this Figure, all non-SM masses are chosen equal to $\msusy$. We compare the results of the different pole determination methods for two different scale choices: $\mu_R = M_t$ (solid), the default choice in \FH, and $\mu_R = \msusy$ (dashed). We set $\tan\beta = 4$. Looking at the curves produced using the numerical pole determination with \DR Higgs field renormalization (blue), we see a small discrepancy between the two scale choices already for low SUSY scales. For rising $\msusy$, the discrepancy increases to up to 12~GeV for $\msusy \sim 100 \tev$. In contrast, the curves using the fixed-order pole determination (red) lie within 1~GeV from each other indicating a small dependence on $\mu_R$ over the whole considered $\msusy$ range. This remaining dependence can be explained by the renormalization scale dependence of $\tan\beta$. Here, the scale dependence of $\mu$ only has a very small impact on the final result for $M_h$. This is confirmed by the result employing numerical pole determination with heavy-OS Higgs field renormalization, where $\tan\beta$ is defined at the fixed scale $Q_\text{pd} = M_t$. The corresponding curves (green) lie on top of each other. Note that also here in principle higher order terms including a dependence on the renormalization scale are induced. As explained in \Sec{subsec:hOS_poledet}, this renofrmalization scale dependence is, however, absorbed into the field renormalization constants and thereby replaced by a dependence on the pole determination scale $Q_\text{pd}$, which is kept fixed. The fact that the results using different renormalization scale choices are equal to each other confirms that this scale separation works as expected and also that the scale dependence of $\mu$ is negligible in the discussed scenario.

These plots show that using numerical pole determination with heavy-OS Higgs field renormalization allows to independently control the different scales of the calculation, i.e., the scales of the pole determination and the scale of $\tan\beta$. This is an important step towards an improved estimate of the remaining theoretical uncertainties.


\section{Conclusions}
\label{sec:06_conclusion}

In this paper we investigated the determination of the Higgs-boson propagator poles in the MSSM. First, we gave a short overview of the MSSM Higgs sector and its renormalization. We especially focused on the Higgs field renormalization.

Afterwards, we discussed the behavior of general $n$-point vertex functions if the masses of one or more internal particles approach infinity. These contributions have a dependence on the external momenta only if the number of external particles is equal two. In this case, the normalization of the external fields can be used to absorb the dependence.

Next, we focused on the determination of the Higgs-boson propagator poles in the MSSM. We showed that the most direct approach, determining the poles numerically employing a \DR Higgs field renormalization, induces higher order terms which would cancel in a more complete calculation. The origin of these terms is the momentum dependence of diagrams involving heavy internal particles.

As an alternative approach to avoid this problem, we introduced a method to determine the Higgs-boson propagator poles at a fixed order. This ensures that no unwanted higher order terms are induced. Close to crossing points, where two Higgs bosons that mix with each other are almost mass degenerate, this methods, however, suffers from unphysical discontinuities in the mass prediction.

Therefore, we introduced a third method which employs a numerical pole determination but fixes the Higgs field renormalization such that the momentum dependence of heavy contributions is absorbed. We call this scheme heavy-OS Higgs field renormalization. This method avoids the inclusion of unwanted higher order terms and also does not suffer from unphysical discontinuities close to crossings points.

In our numerical investigation, we compared the different methods in scenarios with light non-SM Higgs bosons finding differences of up to $\sim 3\gev$ for $\msusy \sim 10\tev$ and ${\tan\beta\sim 1}$. In addition, we closely investigated the predictions for the $M_H^{125}$ benchmark scenario introduced in~\cite{Bahl:2018zmf}, finding the third method to yield a smooth prediction approaching the results of the fixed-order pole determination away from the crossing point. Moreover, we found that the numerical pole determination with heavy-OS Higgs field renormalization brings the prediction of the combined fixed-order and EFT calculation closer to the prediction of the pure EFT calculation for very high SUSY scales. Last, we showed that the complete dependence of the overall result on the renormalization scale is caused by higher order terms induced through the pole determination. This is an important input for a re-evaluation of the remaining theoretical uncertainty, which we leave for future work.

We want to point out that despite the discussion being very specific to the MSSM, the argumentation and the methods presented can straightforwardly be applied to the determination of propagator poles in other models which involve a large mass hierarchy.


\section*{Acknowledgments}
\sloppy{
We thank Wolfgang Hollik, Sven Heinemeyer, Georg Weiglein for helpful comments and carefully reading the manuscript. We thank Thomas Hahn, Stefan Liebler, Sebastian Paßehr, Heidi Rzehak, Pietro Slavich, Ivan Sobolev as well as Tim Stefaniak for useful discussions.
}


\appendix


\section{One- and two-loop counterterms}
\label{app:07_ren}

In this Appendix, we list explicit expressions needed for the renormalization of the MSSM Higgs sector as described in \Sec{sec:02_MSSMintro}. The expressions extend the ones listed e.g.~in \cite{Frank:2006yh,Hollik:2014bua,Bahl:2018jom} by including off-diagonal field renormalization constants (see \Sec{subsubsec:fieldren}) and independent counterterms for $\tan\beta$ and the mixing angles (see discussion in \App{app:09_finfieldren}).

\subsection{One-loop renormalization}

The one-loop field renormalization constants of the neutral Higgs bosons in the mass basis are given by
\begin{subequations}
\begin{align}
\dZOL{hh} &= \saa \dZOL{11} - \sae\cae \left(\dZOL{12} + \dZOL{21}\right) + \caa \dZOL{22}, \\
\dZOL{HH} &= \caa \dZOL{11} + \sae\cae \left(\dZOL{12} + \dZOL{21}\right) + \saa \dZOL{22}, \\
\dZOL{hH} &= \sae\cae \left(\dZOL{22} - \dZOL{11}\right) + \frac{1}{2}c_{2\alpha}\left(\dZOL{12}+\dZOL{21}\right), \\
\dZOL{AA} &= \sbnbn \dZOL{11} - \sbne\cbne \left(\dZOL{12} + \dZOL{21}\right) + \cbnbn \dZOL{22}, \\
\dZOL{GG} &= \cbnbn \dZOL{11} + \sbne\cbne \left(\dZOL{12} + \dZOL{21}\right) + \sbnbn \dZOL{22}, \\
\dZOL{AG} &= \sbne\cbne \left(\dZOL{22} - \dZOL{11}\right) + \frac{1}{2}c_{2\beta_n}\left(\dZOL{12}+\dZOL{21}\right), \\
\dZOL{hA} &= \frac{i}{2}s_{\beta_n-\alpha}\left(\dZOL{12}-\dZOL{21}\right), \\
\dZOL{hG} &= -\frac{i}{2}c_{\beta_n-\alpha}\left(\dZOL{12}-\dZOL{21}\right), \\
\dZOL{HA} &= \frac{i}{2}c_{\beta_n-\alpha}\left(\dZOL{12}-\dZOL{21}\right), \\
\dZOL{HG} &= \frac{i}{2}s_{\beta_n-\alpha}\left(\dZOL{12}-\dZOL{21}\right).
\end{align}
\end{subequations}
The corresponding constants in the charged Higgs sector read
\begin{subequations}
\begin{align}
\dZOL{\Hpm\Hpm} &= \sbcbc \dZOL{11} - \sbce\cbce \left(\dZOL{12} + \dZOL{21}\right) + \cbcbc \dZOL{22}, \\
\dZOL{\Gpm\Gpm} &= \cbcbc \dZOL{11} + \sbce\cbce \left(\dZOL{12} + \dZOL{21}\right) + \sbcbc \dZOL{22}, \\
\dZOL{\Hpm\Gpm} &= \sbce\cbce \left(\dZOL{22} - \dZOL{11}\right) + \cbcbc\dZOL{21} - \sbcbc\dZOL{12}, \\
\dZOL{\Gpm\Hpm} &= \sbce\cbce \left(\dZOL{22} - \dZOL{11}\right) + \cbcbc\dZOL{12} - \sbcbc\dZOL{21}.
\end{align}
\end{subequations}
The field renormalization counterterms in the gaugeless limit can be obtained by setting $\alpha = \beta - \pi/2$ in the expressions listed above.

The one-loop mass counterterms of the neutral Higgs sector are given by
\begin{subequations}
\begin{align}
\dOL m_h^2 ={}& \dOL M_A^2 c_{\alpha-\beta}^2 + \dOL M_Z^2 s_{\alpha+\beta}^2 \nonumber\\
&+ \frac{e}{2 M_W s_W c_W}\left(\dOL T_H c_{\alpha-\beta}s_{\alpha-\beta}^2 + \dOL T_h s_{\alpha-\beta}(1+c_{\alpha-\beta}^2)\right) \nonumber\\
&+ \cbb\dOL\tbe (M_Z^2 s_{2(\alpha+\beta)} + M_A^2 s_{2(\alpha-\beta)}) \nonumber\\
&+ \caa\dOL\tae (M_Z^2 s_{2(\alpha+\beta)}-M_A^2 s_{2(\alpha-\beta)}), \\
\dOL m_{hH}^2 ={}& \frac{1}{2}\left(\dOL M_A^2 s_{2(\alpha-\beta)} - \dOL M_Z^2 s_{2(\alpha+\beta)}\right) \nonumber\\
&+ \frac{e}{2 M_W s_W c_W}\left(\dOL T_H s_{\alpha-\beta}^3 - \dOL T_h c_{\alpha-\beta}^3\right) \nonumber\\
&- \cbb\dOL\tbe (M_Z^2 c_{2(\alpha+\beta)} + M_A^2 c_{2(\alpha-\beta)}) \nonumber\\
&- \caa\dOL\tae (M_Z^2 c_{2(\alpha+\beta)}-M_A^2 c_{2(\alpha-\beta)}), \label{eq:dmhH}\\
\dOL m_{H}^2 ={}& \dOL M_A^2 s_{\alpha-\beta}^2 + \dOL M_Z^2 c_{\alpha+\beta}^2 \nonumber\\
&- \frac{e}{2 M_W s_W c_W}\left(\dOL T_H c_{\alpha-\beta}(1+s_{\alpha-\beta}^2) + \dOL T_h s_{\alpha-\beta}c_{\alpha-\beta}^2\right) \nonumber\\
&- \cbb\dOL\tbe (M_Z^2 s_{2(\alpha+\beta)} + M_A^2 s_{2(\alpha-\beta)}) \nonumber\\
&- \caa\dOL\tae (M_Z^2 s_{2(\alpha+\beta)}-M_A^2 s_{2(\alpha-\beta)}),\\
\dOL m_{AG}^2 ={}& \frac{e}{2 M_W s_W c_W}\left(-\dOL T_H s_{\alpha-\beta} - \dOL T_h c_{\alpha-\beta}\right) \nonumber\\
&- M_A^2 \cbb\left(\dOL\tbe - \dOL\tbne\right),\\
\dOL m_{G}^2 ={}& \frac{e}{2 M_W s_W c_W}\left(-\dOL T_H c_{\alpha-\beta} + \dOL T_h s_{\alpha-\beta}\right)\\
\dOL m_{hA}^2 ={}& \frac{e}{2 M_W s_W c_W}\dOL T_A s_{\alpha-\beta},\\
\dOL m_{HA}^2 ={}& - \dOL m_{hG}^2, \\
\dOL m_{HG}^2 ={}& \dOL m_{hA}^2
\end{align}
\end{subequations}
and for the charged Higgs sector by
\begin{subequations}
\begin{align}
\dOL m_{\Hpm}^2 ={}& \dOL M_A^2 - \dOL M_W^2, \\
\dOL m_{\Hpm\Gpm}^2 ={}& \frac{e}{2 M_W s_W c_W}\left(-\dOL T_H s_{\alpha-\beta} - \dOL T_h c_{\alpha-\beta} - i \dOL T_A \right) \nonumber\\
&- m_{H^\pm}^2 \cbb \left(\dOL\tbe - \dOL\tbce\right) \\
\dOL m_{\Gpm\Hpm}^2 ={}& \left(\dOL m_{H^\pm G^\pm}^2\right)^* \\
\dOL m_{\Gpm}^2 ={}& \frac{e}{2 M_W s_W c_W}\left(-\dOL T_H c_{\alpha-\beta} + \dOL T_h s_{\alpha-\beta}\right).
\end{align}
\end{subequations}


\subsection{Two-loop renormalization}

The relations between the two-loop field renormalization constants of the neutral Higgs sector in the mass basis and in the original gauge basis are given by
\begin{subequations}
\begin{align}
\dZTL{hh} &= \saa \DZ{11} - \sae\cae \left(\DZ{12} + \DZ{21}\right) + \caa \DZ{22}, \\
\dZTL{HH} &= \caa \DZ{11} + \sae\cae \left(\DZ{12} + \DZ{21}\right) + \saa \DZ{22}, \\
\dZTL{hH} &= \sae\cae \left(\DZ{22} - \DZ{11}\right) + \frac{1}{2}c_{2\alpha}\left(\DZ{12}+\DZ{21}\right), \\
\dZTL{AA} &= \sbnbn \DZ{11} - \sbne\cbne \left(\DZ{12} + \DZ{21}\right) + \cbnbn \DZ{22}, \\
\dZTL{AG} &= \sbne\cbne \left(\DZ{22} - \DZ{11}\right) + \frac{1}{2}c_{2\beta_n}\left(\DZ{12}+\DZ{21}\right), \\
\dZTL{hA} &= \frac{i}{2}s_{\beta_n-\alpha}\left(\DZ{12}-\DZ{21}\right), \\
\dZTL{HA} &= \frac{i}{2}c_{\beta_n-\alpha}\left(\DZ{12}-\DZ{21}\right),
\end{align}
\end{subequations}
where $\DZ{ij} = \dZTL{ij} - \frac{1}{4}(\dZOL{ij})^2$. The corresponding expressions for the charged Higgs sector read
\begin{subequations}
\begin{align}
\dZTL{\Hpm\Hpm} &= \sbcbc \DZ{11} - \sbce\cbce \left(\DZ{12} + \DZ{21}\right) + \cbcbc \DZ{22}, \\
\dZTL{\Gpm\Gpm} &= \cbcbc \DZ{11} + \sbce\cbce \left(\DZ{12} + \DZ{21}\right) + \sbcbc \DZ{22}, \\
\dZTL{\Hpm\Gpm} &= \sbce\cbce \left(\DZ{22} - \DZ{11}\right) + \cbcbc\DZ{21} - \sbcbc\DZ{12}, \\
\dZTL{\Gpm\Hpm} &= \sbce\cbce \left(\DZ{22} - \DZ{11}\right) + \cbcbc\DZ{12} - \sbcbc\DZ{21}.
\end{align}
\end{subequations}
The field renormalization counterterms in the gaugeless limit can be obtained by setting $\alpha = \beta - \pi/2$.

The two-loop mass counterterms in the neutral Higgs sector are given by
\begin{subequations}
\begin{align}
\dTL m_h^2 ={}& M_A^2 \left(\cbb\dOL\tbe - \sbb\dOL\tae\right)^2 \nonumber\\
&- \frac{e}{2 M_W s_W}\left[\delta^{(2)}T_h + \dOL T_h\delta^{(1)}Z_W + \dOL T_H\left(\cbb\dOL\tbe - \sbb\dOL\tae\right) \right],\\
\dTL m_{hH}^2 ={}& M_A^2\left[\cbb\dTL\tbe - \sbb\dTL\tae - \sbe\cbe^3(\dOL\tbe)^2 - \sbe^3\cbe(\dOL\tae)^2\right] \nonumber\\
& + \dOL M_A^2\left(\cbb\dOL\tbe - \sbb\dOL\tae\right) \nonumber\\
&- \frac{e}{2 M_W s_W}\left[\delta^{(2)}T_H+\delta^{(1)}T_H\delta^{(1)}Z_W \right],\\
\dTL m_H^2 =& \delta^{(2)}M_A^2 + \frac{e}{M_W s_W}\dOL T_H \left(\cbb\dOL\tbne - \sbb\dOL\tae\right) \nonumber\\
&+ M_A^2\left[\cbe^4(\dOL\tbne)^2 - \sbe^4(\dOL\tae)^2 - 2\cbb\dOL\tbe\left(\cbb\dOL\tbne - \sbb\dOL\tae\right)\right],\\
\dTL m_{A G}^2 ={}& M_A^2\cbb \left[\dTL\tbne - \dTL\tbe + \sbe\cbe\left((\dOL\tbe)^2-(\dOL\tbne)^2\right)\right] \nonumber\\
& + \dOL M_A^2\cbb\left(\dOL\tbne - \dOL\tbe\right) \nonumber\\
& + \frac{e}{2 M_W s_W}\left[\dTL T_H + \dOL T_H\delta^{(1)}Z_W + \dOL T_h \left(\cbb\dOL\tbne - \sbb\dOL\tae\right) \right],\\
\dTL m_G^2 ={}& M_A^2\cbe^4 \left(\dOL\tbne - \dOL\tbe\right)^2 \nonumber\\
& - \frac{e}{2 M_W s_W}\left[\dTL T_h  + \dOL T_h\delta^{(1)}Z_W  \right.\nonumber\\
&\left.\hspace{2.3cm}- \dOL T_H \left(2\cbb\dOL\tbne - \cbb\dOL\tbe - \sbb\dOL\tae\right) \right],\\
\dTL m_{hA}^2 ={}& - \frac{e}{2 M_W s_W} \left(\dTL T_A  + \dOL T_A\delta^{(1)}Z_W\right), \\
\dTL m_{hG}^2 ={}& - \frac{e}{2 M_W s_W} \dOL T_A \left(\cbb\dOL\tbne - \sbb\dOL\tae\right), \\
\dTL m_{HA}^2 ={}& - \dTL m_{hG}^2, \\
\dTL m_{HG}^2 ={}& \dTL m_{hA}^2
\end{align}
\end{subequations}
and for the charged Higgs sector by
\begin{subequations}
\begin{align}
\dTL m_{\Hpm\Gpm}^2 ={}& m_{H^\pm}^2\cbb \left[\dTL\tbce - \dTL\tbe + \sbe\cbe\left((\dOL\tbe)^2-(\dOL\tbce)^2\right)\right] \nonumber\\
& + \dOL m_{H^\pm}^2\cbb\left(\dOL\tbce - \dOL\tbe\right) \nonumber\\
& + \frac{e}{2 M_W s_W}\left[\dTL T_H - i \dTL T_A + \dOL T_h \left(\cbb\dOL\tbce - \sbb\dOL\tae\right) \right],\\
\dTL m_{\Gpm\Hpm}^2 ={}& (\dTL m_{H^\pm G^\pm}^2)^*,\\
\dTL m_{\Gpm}^2 ={}& m_{H^\pm}^2\cbe^4 \left(\dOL\tbce - \dOL\tbe\right)^2 \nonumber\\
& - \frac{e}{2 M_W s_W}\left[\dTL T_h - \dOL T_H \left(2\cbb\dOL\tbce - \cbb\dOL\tbe - \sbb\dOL\tae\right) \right].
\end{align}
\end{subequations}
Note that the expressions are only valid in the limit of vanishing electroweak gauge couplings. Correspondingly, also the counterterm of $\alpha$ has to be inserted applying the limit of vanishing electroweak gauge couplings.

The remaining two-loop mass counterterms, $\dmTL{A}$ and $\dmTL{\Hpm}$, are fixed by employing OS conditions for either the mass of the $A$ boson or the mass of the $\Hpm$ boson (and by using $\dmTL{A} = \dmTL{\Hpm}$, which holds also only in the limit of vanishing electroweak gauge couplings).

In order to obtain the renormalized two-loop self-energies (see \Eq{eq:renTLse}) the following combinations of mass and field counterterms are needed in the neutral Higgs sector
\begin{subequations}\label{Eq:TLseCTs}
\begin{align}
\delta^{(2)}m_h^{\textbf{Z}} ={}
&   \frac{1}{4}M_A^2\left[\left(\dZOL{hH}\right)^2+\left(\dZOL{hA}\right)^2\right] \nonumber\\
& + \dZOL{hh}\dmOL{h} + \dZOL{hH}\dmOL{hH} \nonumber\\
& + \dZOL{hA}\dmOL{hA} + \dZOL{hG}\dmOL{hG} \nonumber\\
& + \dmTL{h}, \\
\delta^{(2)}m_H^{\textbf{Z}} ={}
&   M_A^2\left[\dZTL{HH}+\frac{1}{4}\left(\dZOL{HH}\right)^2+\frac{1}{4}\left(\dZOL{HA}\right)^2\right]\nonumber\\
& + \dZOL{HH}\dmOL{H}+\dZOL{hH}\dmOL{hH} \nonumber\\
& + \dZOL{HA}\dmOL{HA}+\dZOL{HG}\dmOL{HG} \nonumber\\
& + \dmTL{H},\\
\delta^{(2)}m_A^{\textbf{Z}} ={}
&   M_A^2\left[\dZTL{AA}+\frac{1}{4}\left(\dZOL{AA}\right)^2+\frac{1}{4}\left(\dZOL{HA}\right)^2\right]\nonumber\\
& + \dZOL{AA}\dmOL{A}+\dZOL{AG}\dmOL{AG} \nonumber\\
& + \dZOL{hA}\dmOL{hA}+\dZOL{HA}\dmOL{HA} \nonumber\\
& + \dmTL{A},\\
\delta^{(2)}m_{hH}^{\textbf{Z}} ={}
&   \frac{1}{2}M_A^2\left[\delta^{(2)}Z_{hH} + \frac{1}{2}\dZOL{HH}\dZOL{hH} + \frac{1}{2}\dZOL{hA}\dZOL{HA}\right] \nonumber\\
& + \frac{1}{2}\left[\left(\dZOL{hh}+\dZOL{HH}\right)\dmOL{hH}+\dZOL{hH}\left(\dmOL{h}+\dmOL{H}\right)\right] \nonumber\\
& + \frac{1}{2}\left[\dZOL{hA}\dmOL{HA}+\dZOL{HA}\dmOL{hA}+\dZOL{hG}\dmOL{HG}+\dZOL{HG}\dmOL{hG}\right] \nonumber\\
& + \dmTL{hH},\\
\delta^{(2)}m_{hA}^{\textbf{Z}} ={}
&   \frac{1}{2}M_A^2\left[\delta^{(2)}Z_{hA} + \frac{1}{2}\dZOL{hA}\dZOL{AA} + \frac{1}{2}\dZOL{hH}\dZOL{HA}\right] \nonumber\\
& + \frac{1}{2}\left[\left(\dZOL{hh}+\dZOL{AA}\right)\dmOL{hA}+\dZOL{hH}\dmOL{HA}+\dZOL{AG}\dmOL{hG}\right] \nonumber\\
& + \frac{1}{2}\left[\dZOL{hA}\left(\dmOL{hh}+\dmOL{AA}\right)+\dZOL{hG}\dmOL{AG}+\dZOL{HA}\dmOL{hH}\right] \nonumber\\
& + \dmTL{hA},\\
\delta^{(2)}m_{HA}^{\textbf{Z}} ={}
&   M_A^2\left[\delta^{(2)}Z_{HA} + \frac{1}{4}\dZOL{HH}\dZOL{HA} + \frac{1}{4}\dZOL{HA}\dZOL{AA}\right] \nonumber\\
& + \frac{1}{2}\left[\left(\dZOL{HH}+\dZOL{AA}\right)\dmOL{HA}+\dZOL{hH}\dmOL{hA}+\dZOL{AG}\dmOL{HG}\right] \nonumber\\
& + \frac{1}{2}\left[\dZOL{HA}\left(\dmOL{HH}+\dmOL{AA}\right)+\dZOL{HG}\dmOL{AG}+\dZOL{hA}\dmOL{hH}\right] \nonumber\\
& + \dmTL{HA}
\end{align}
\end{subequations}
and the charged Higgs sector
\begin{align}
\delta^{(2)}m_{\Hpm}^{\textbf{Z}} ={}
& m_{\Hpm}^2\left[\dZTL{\Hpm\Hpm}+\frac{1}{4}\left(\dZOL{\Hpm\Hpm}\right)^2\right] \nonumber\\
& + \dZOL{\Hpm\Hpm}\dmOL{\Hpm} \nonumber\\
& + \frac{1}{2}\left(\dZOL{\Hpm\Gpm}\dmOL{\Gpm\Hpm}+\dZOL{\Gpm\Hpm}\dmOL{\Hpm\Gpm}\right)\nonumber\\
& + \dmTL{\Hpm}.
\end{align}
These expressions are again only valid in the gaugeless limit. Correspondingly, this limit has also to be applied for the appearing field renormalization constants.

The two-loop tadpole counterterms are obtained by
\begin{samepage}
\begin{subequations}
\begin{align}
\dTL T_h ={}& - \frac{1}{2}\left(\delta^{(1)}Z_{hh}\delta^{(1)}T_h+\delta^{(1)}Z_{hH}\delta^{(1)}T_H \right.\nonumber\\
&\left.\hspace{.9cm}+\delta^{(1)}Z_{hA}\delta^{(1)}T_A+\delta^{(1)}Z_{hG}\delta^{(1)}T_G\right), \\
\dTL T_H ={}& -\frac{1}{2}\left(\delta^{(1)}Z_{HH}\delta^{(1)}T_H+\delta^{(1)}Z_{hH}\delta^{(1)}T_h \right.\nonumber\\
&\left.\hspace{.9cm}+\delta^{(1)}Z_{HA}\delta^{(1)}T_A+\delta^{(1)}Z_{HG}\delta^{(1)}T_G\right), \\
\dTL T_A ={}& - \frac{1}{2}\left(\delta^{(1)}Z_{AA}\delta^{(1)}T_A+\delta^{(1)}Z_{AG}\delta^{(1)}T_G \right.\nonumber\\
&\left.\hspace{.9cm}+\delta^{(1)}Z_{hA}\delta^{(1)}T_h+\delta^{(1)}Z_{HA}\delta^{(1)}T_H\right).
\end{align}
\end{subequations}
\end{samepage}

\clearpage
\section{Reparametrization of \texorpdfstring{$\boldsymbol{\tan}\boldsymbol{\beta}$}{tanb}}
\label{app:08_tbrepara}

The heavy-OS Higgs field renormalization fixes $\tan\beta$ to be defined in the THDM at the scale $M_t$ (see discussion in \Sec{sec:04_poledet}). Here, we discuss how a different definition for $\tan\beta$ can be achieved.

The most direct solution is to accept that $\tan\beta$ is defined as stated above. If the input $\tan\beta$ is then defined in another scheme, we have to convert it to the scheme of the calculation,
\begin{align}
\tbe^\text{THDM}(M_t) = \tbe^\text{in}(Q_\text{in}) + \Delta\tbe \; ,
\end{align}
where $\Delta\tbe$ is a generic loop correction.

The obtained value is then inserted as new input parameter into the Higgs mass calculation. In practice, a one-loop conversion is employed. This procedure is easy. It, however, has the disadvantage that the conversion induces terms beyond the order of the fixed-order conversion which should not be included if the result of the fixed-order calculation is combined with an infinite series of resummed logarithms.\footnote{This issue is similar to problems with the conversion of the stop mixing parameter from the \DR to the OS scheme employed in earlier \FH versions (see \cite{Bahl:2017aev} for more details).}

To avoid the problem with induced higher-order terms, we can expand the result e.g.\ for $M_h^2$ around the input $\tan\beta$,
\begin{align}\label{eq:tanb_expansion}
M_h^2(\tbe^\text{in}(Q_\text{in})) = M_h^2(\tbe^\text{THDM}(M_t)) - \frac{\partial}{\partial\tan\beta}M_h^2(\tbe^\text{THDM}(M_t)) \cdot \Delta\tbe + \ldots,
\end{align}
where the expansion is truncated at the same order as the fixed-order calculation.

This approach can be implemented by introducing an additional counterterm for $\tan\beta$ which is independent of the Higgs field renormalization. The $\tan\beta$ counterterms are then given by
\begin{subequations}
\begin{align}
\dOL\tbe ={}& \frac{1}{2}\tbe\left(\dZOL{22} - \dZOL{11}\right) + \frac{1}{4}\left(1-\tbb)(\dZOL{12}+\dZOL{21}\right) + \dOL \tbe \big|_\text{free}\\
\dTL\tbe\gl ={}&\bigg[
                 \frac{1}{2}\tbe\left(\dZTL{22}-\dZTL{11}\right) +\frac{1}{4}\left(1-\tbb\right)\left(\dZTL{12}+\dZTL{21}\right) \nonumber\\
&\hspace{.1cm} - \frac{1}{2}\tbe\left(\dZOL{22}-\dZOL{11}\right)\left(\frac{1}{4}\left(\dZOL{22}-\dZOL{11}\right)+\dZOL{11}\right) \nonumber\\
&\hspace{.1cm} - \frac{1}{8}\left((1-2\tbb)\dZOL{11} + \tbb\dZOL{22}\right)\left(\dZOL{12}+\dZOL{21}\right) \nonumber\\
&\hspace{.1cm} - \frac{1}{16}(1-\tbb)\left((1+\tbe)\left((\dZOL{12})^2 + (\dZOL{21})^2\right) + 2 \tbe\dZOL{12}\dZOL{21}\right)
\bigg]_\text{gl} \nonumber \\
&+ \dTL \tbe \big|_\text{free,gl}.
\end{align}
\end{subequations}
The additional terms, labelled by the subscript ``free'', can be chosen independent of the Higgs field renormalization.\footnote{This corresponds to allowing for finite non-zero contributions from the the vev counterterms in~\Eq{eq:vevren}.} In order to achieve $\tan\beta$ to be defined e.g.\ at the renormalization scale of the fixed-order calculation in the MSSM, we have to chose these terms such that
\begin{align}
\dOL\tbe\big|_\text{fin} &= 0,\\
\dTL\tbe\big|_\text{gl,fin} &= 0.
\end{align}
In addition to adding additional counterterms for $\tan\beta$, we also add counterterms for the Higgs mixing angles. These are normally not renormalized. They, however, intrinsically depend on $\tan\beta$. Therefore, the expansion in \Eq{eq:tanb_expansion} generates terms that can most easily be taken into account by introducing finite counterterms for the mixing angles,
\begin{subequations}
\begin{align}
t_{\alpha } &\rightarrow t_{\alpha } + \dOL t_{\alpha } + \dTL t_{\alpha },\\
t_{\beta_n} &\rightarrow t_{\beta_n} + \dOL t_{\beta_n} + \dTL t_{\beta_n},\\
t_{\beta_c} &\rightarrow t_{\beta_c} + \dOL t_{\beta_c} + \dTL t_{\beta_c}.
\end{align}
\end{subequations}
At the one-loop level, the following relations between the counterterms of $\tan\beta$ and the mixing angles hold,
\begin{subequations}
\begin{align}
\dOL t_{\alpha } &= \frac{2(M_A^4 - M_Z^4)\cbb}{(m_H^2-m_h^2)(m_H^2-m_h^2 - (M_A^2 - M_Z^2)c_{2\beta})}\; \dOL\tbe\big|_\text{free}, \\
\dOL t_{\beta_n} &= \dOL\tbe\big|_\text{free}, \\
\dOL t_{\beta_c} &= \dOL\tbe\big|_\text{free},
\end{align}
\end{subequations}
and at the two-loop level,
\begin{subequations}
\begin{align}
\dTL\tae\big|_\text{gl} &= \frac{1}{\tbb}\left(\dTL\tbe\big|_\text{free,gl}-\frac{1}{\tbe}(\dOL\tbe\big|_\text{free,gl})^2\right),\\
\dTL\tbne\big|_\text{gl} = \dTL\tbce\big|_\text{gl} &= \dTL\tbe\big|_\text{free,gl}.
\end{align}
\end{subequations}
We find that both methods yield numerically very similar results. Only in scenarios with low $\tan\beta$ and a large hierarchy between $Q_\text{in}$ and $M_t$, we observe non-negligible, but still small, numerical differences (e.g.~for $Q_\text{in}= \msusy = 20\tev$ and $\tan\beta\sim 2$ the difference in $M_h$ is $\lesssim 0.4\gev$). These are induced through the higher-order terms generated in case of a conversion of the input $\tan\beta$.


\section{Heavy-OS renormalization - explicit expressions}
\label{app:09_finfieldren}

Here, we list explicit expressions for the finite parts of the Higgs field renormalization constants in the heavy-OS scheme. Similar expressions have already been presented in \cite{Bahl:2018jom}. In there, however, all parameters were assumed to be real and all SUSY soft-breaking scalar masses as well as the gaugino masses were assumed to equal.

For abbreviation, we introduce the constant
\begin{align}
c = -\frac{\alpha}{16\pi M_W^2 s_W^2},
\end{align}
where $\alpha$ is the fine structure constant and $s_W$ is the sine of the electroweak mixing angle.

The contribution from up-type squarks is given by
\begin{subequations}
\begin{align}
\dZOL{11}\Big|_{\tilde u, \tilde c, \tilde t} &= + 6 c \sum_{f = u, c, t} \frac{m_f^2}{\sbb}\; \mu\mu^* \; B_0'\left(0, m_{\tilde Q_f}^2, m_{\tilde U_f}^2\right),\\
\dZOL{12}\Big|_{\tilde u, \tilde c, \tilde t} &= - 6 c \sum_{f = u, c, t} \frac{m_f^2}{\sbb}\; A_f\mu \; B_0'\left(0, m_{\tilde Q_f}^2, m_{\tilde U_f}^2\right),\\
\dZOL{22}\Big|_{\tilde u, \tilde c, \tilde t} &= + 6 c \sum_{f = u, c, t} \frac{m_f^2}{\sbb}\; A_f A_f^* \; B_0'\left(0, m_{\tilde Q_f}^2, m_{\tilde U_f}^2\right).
\end{align}
\end{subequations}
$B_0'$ marks the derivative of the corresponding one-loop Passarino-Veltman loop function with respect to the external momentum. $m_{\tilde Q_f}$ denotes the SUSY-soft breaking mass for left-handed squarks. $m_{\tilde Q_f}$ is the SUSY-soft breaking mass for right-handed up-type squarks. $A_f$ denotes the soft-breaking trilinear coupling.

The corresponding contributions from down-type squarks read
\begin{subequations}
\begin{align}
\dZOL{11}\Big|_{\tilde d, \tilde s, \tilde b} &= + 6 c \sum_{f = d, s, b} \frac{m_f^2}{\cbb}\; A_fA_f^* \; B_0'\left(0, m_{\tilde Q_f}^2, m_{\tilde D_f}^2\right),\\
\dZOL{12}\Big|_{\tilde d, \tilde s, \tilde b} &= - 6 c \sum_{f = d, s, b} \frac{m_f^2}{\cbb}\; A_f\mu \; B_0'\left(0, m_{\tilde Q_f}^2, m_{\tilde D_f}^2\right),\\
\dZOL{22}\Big|_{\tilde d, \tilde s, \tilde b} &= + 6 c \sum_{f = d, s, b} \frac{m_f^2}{\cbb}\; \mu\mu^* \; B_0'\left(0, m_{\tilde Q_f}^2, m_{\tilde D_f}^2\right)
\end{align}
\end{subequations}
Here, $m_{\tilde D_f}$ is the SUSY-soft breaking mass for right-handed down-type squarks.

The slepton contribution reads
\begin{subequations}
\begin{align}
\dZOL{11}\Big|_{\tilde e, \tilde \mu, \tilde \tau} &= + 2 c \sum_{l = e, \mu, \tau} \frac{m_l^2}{\cbb}\; A_lA_l^* \; B_0'\left(0, m_{\tilde L_l}^2, m_{\tilde E_l}^2\right),\\
\dZOL{12}\Big|_{\tilde e, \tilde \mu, \tilde \tau} &= - 2 c \sum_{l = e, \mu, \tau} \frac{m_l^2}{\cbb}\; A_l\mu \; B_0'\left(0, m_{\tilde L_l}^2, m_{\tilde E_l}^2\right),\\
\dZOL{22}\Big|_{\tilde e, \tilde \mu, \tilde \tau} &= + 2 c \sum_{l = e, \mu, \tau} \frac{m_l^2}{\cbb}\; \mu\mu^* \; B_0'\left(0, m_{\tilde L_l}^2, m_{\tilde E_l}^2\right),
\end{align}
\end{subequations}
where $m_{\tilde L_l}$ is the SUSY-soft breaking mass for right-handed sleptons. $m_{\tilde E_l}$ is the SUSY-soft breaking parameter for left-handed sleptons.

The contribution from electroweakinos (superpartners of the electroweak gauge bosons and the Higgs bosons) are obtained via
\begin{samepage}
\begin{subequations}
\begin{align}
\dZOL{11}\Big|_\text{EWino} ={}& -4 c\; M_Z^2 \left\{s_W^2 \left[B_1\left(0, |M_1|^2, |\mu|^2\right) + |M_1|^2 B_0'\left(0, |M_1|^2, |\mu|^2\right)\right] \right.\nonumber\\
&\left.\hspace{1.9cm}+ c_W^2 \left[B_1\left(0, |M_2|^2, |\mu|^2\right) - B_0\left(0, |M_2|^2, |\mu|^2\right) \right.\right.\nonumber\\
&\left.\left.\hspace{3cm}+ (2|M_2|^2 + |\mu|^2)B_0'\left(0, |M_2|^2, |\mu|^2\right)\right]\right\} ,\\
\dZOL{12}\Big|_\text{EWino} ={}& - 4 c\; M_Z^2 |\mu|\left\{s_W^2 |M_1|B_0'\left(0, |M_1|^2, |\mu|^2\right) \right.\nonumber\\
&\left.\hspace{2.4cm}+ 3 c_W^2 |M_2| B_0'\left(0, |M_2|^2, |\mu|^2\right)\right\},\\
\dZOL{22}\Big|_\text{EWino} ={}& \dZOL{11}\Big|_\text{EWino},
\end{align}
\end{subequations}
\end{samepage}\noindent
where $M_1$ and $M_2$ are the SUSY soft-breaking gaugino masses. $c_W$ is the cosine of the electroweak mixing angle.

In addition, there is a contribution from non-SM Higgs bosons,
\begin{samepage}
\begin{subequations}
\begin{align}
\dZOL{11}\Big|_\text{heavyH} ={}& - c (2 M_W^2 + M_Z^2)\left(3 - 2 \ln(M_A^2/\mu_R^2)\right)\sbb ,\\
\dZOL{12}\Big|_\text{heavyH} ={}& + c (2 M_W^2 + M_Z^2)\left(3 - 2 \ln(M_A^2/\mu_R^2)\right)\sbe\cbe,\\
\dZOL{22}\Big|_\text{heavyH} ={}& - c (2 M_W^2 + M_Z^2)\left(3 - 2 \ln(M_A^2/\mu_R^2)\right)\cbb.
\end{align}
\end{subequations}
\end{samepage}\noindent
Note that this contribution cancels out in the $\tan\beta$ counterterm. Therefore, it does not affect the definition of $\tan\beta$.

The one-loop ``light'' anomalous dimensions, as defined in \Eq{eq:anomal_dim_light}, gets contributions from SM fermions,
\begin{subequations}
\begin{align}
\gamma^{\light,(1)}_{11}\Big|_{f} &=  2 \frac{c}{\cbb}\left\{3\sum_{f = d,s,b} m_f^2+ \sum_{l = e, \mu, \tau}m_l^2\right\},\\
\gamma^{\light,(1)}_{12}\Big|_{f} &= 0,\\
\gamma^{\light,(1)}_{22}\Big|_{f} &= 6 \frac{c}{\sbb}\sum_{f = u,c,t} m_f^2.
\end{align}
\end{subequations}
and gauge bosons
\begin{subequations}
\begin{align}
\gamma^{\light,(1)}_{11}\Big|_{W,Z} &= - 2 c (2 M_W^2 + M_Z^2)\cbb,\\
\gamma^{\light,(1)}_{12}\Big|_{W,Z} &= - 2 c (2 M_W^2 + M_Z^2)\sbe\cbe,\\
\gamma^{\light,(1)}_{22}\Big|_{W,Z} &= - 2 c (2 M_W^2 + M_Z^2)\sbb.
\end{align}
\end{subequations}


\section{Transformation of the \texorpdfstring{$\boldsymbol{Z}$}{Z}-matrix from the heavy-OS scheme to the \DR scheme}
\label{app:10_Zmatrix}

The $\mathbf{Z}$-matrix relates the tree-level mass eigenstates to the physical mass eigenstates in scattering amplitudes. Restricting us to the case of $2\times 2$ mixing, it is given at the one-loop level by
\begin{align}
\mathbf{Z} =
\begin{pmatrix}
1 - \frac{1}{2} \hat\Sigma_{hh}^{(1)\prime}(m_h^2) & - \frac{\hat\Sigma_{hH}(m_h^2)}{m_h^2 - m_H^2} \\
-\frac{\hat\Sigma_{hH}(m_H^2)}{m_H^2 - m_h^2} & 1 - \frac{1}{2} \hat\Sigma_{HH}^{(1)\prime}(m_H^2)
\end{pmatrix}.
\end{align}
Formulas valid beyond the one-loop level and more details can be found e.g.\ in \cite{Frank:2006yh,Fuchs:2016swt}.

Obviously, the $\mathbf{Z}$-matrix depends explicitly on the choice of the renormalization scheme for the Higgs field renormalization. This explicit dependence can be removed by multiplying with the $hH$ submatrix of the Higgs field renormalization matrix $\mathbf{Z}_{hHAG}$ (see \Eq{eq:ZhHAG}),
\begin{align}
&
\begin{pmatrix}
1 + \frac{1}{2}\dZOL{hh} &     \frac{1}{2}\dZOL{hH} \\
    \frac{1}{2}\dZOL{hH} & 1 + \frac{1}{2}\dZOL{HH}
\end{pmatrix}_\text{fin}
\mathbf{Z} =\nonumber\\
&=
\begin{pmatrix}
1 - \frac{1}{2}\left(\Sigma_{hh}^{(1)\prime}(m_h^2) + \dZOL{hh}\big|_\DR\right) & -\frac{\Sigma_{hH}(m_h^2) - \dmOL{hH}}{m_h^2 - m_H^2} - \frac{1}{2}\dZOL{hH}\big|_\DR \\
-\frac{\Sigma_{hH}(m_H^2) - \dmOL{hH}}{m_H^2 - m_h^2} - \frac{1}{2}\dZOL{hH}\big|_\DR & 1 - \frac{1}{2}\left(\Sigma_{HH}^{(1)\prime}(m_H^2) + \dZOL{HH}\big|_\DR\right)
\end{pmatrix}.
\end{align}
The subscript ``fin'' indicates that only the finite part is taken into account in order to ensure that the transformed $\mathbf{Z}$-matrix is still UV finite.

In addition to the explicit dependence on the Higgs field renormalization constants, there also is an implicit dependence through the $\tan\beta$ counterterm which appears in $\dmOL{hH}$ (see~\Eq{eq:dmhH}). This dependence vanishes if the $\mathbf{Z}$-matrix is combined with renormalized $n$-point vertex functions in order to calculate physical observables.

In order to get a \DR-renormalized $\mathbf{Z}$-matrix, which can easily be used in other calculations, we have to multiply the $\mathbf{Z}$-matrix calculated employing heavy-OS Higgs field renormalization with the corresponding finite part of the Higgs field renormalization matrix. In addition, we have to subtract the pieces proportional to the $\tan\beta$ counterterm in the off-diagonal entries. This procedure is applied in \FH.


\newpage

\bibliographystyle{JHEP}
\bibliography{bibliography}{}

\providecommand{\href}[2]{#2}\begingroup\raggedright\begin{thebibliography}{10}

\bibitem{Aad:2012tfa}
{\scshape ATLAS} collaboration, G.~Aad et~al., \emph{{Observation of a new
  particle in the search for the Standard Model Higgs boson with the ATLAS
  detector at the LHC}},
  \href{http://dx.doi.org/10.1016/j.physletb.2012.08.020}{\emph{Phys. Lett.}
  {\bf B716} (2012) 1--29}, [\href{http://arxiv.org/abs/1207.7214}{{\tt
  1207.7214}}].

\bibitem{Chatrchyan:2012xdj}
{\scshape CMS} collaboration, S.~Chatrchyan et~al., \emph{{Observation of a new
  boson at a mass of 125 GeV with the CMS experiment at the LHC}},
  \href{http://dx.doi.org/10.1016/j.physletb.2012.08.021}{\emph{Phys. Lett.}
  {\bf B716} (2012) 30--61}, [\href{http://arxiv.org/abs/1207.7235}{{\tt
  1207.7235}}].

\bibitem{Aad:2015zhl}
{\scshape ATLAS, CMS} collaboration, G.~Aad et~al., \emph{{Combined Measurement
  of the Higgs Boson Mass in $pp$ Collisions at $\sqrt{s}=7$ and 8 TeV with the
  ATLAS and CMS Experiments}},
  \href{http://dx.doi.org/10.1103/PhysRevLett.114.191803}{\emph{Phys. Rev.
  Lett.} {\bf 114} (2015) 191803}, [\href{http://arxiv.org/abs/1503.07589}{{\tt
  1503.07589}}].

\bibitem{Nilles:1983ge}
H.~P. Nilles, \emph{{Supersymmetry, Supergravity and Particle Physics}},
  \href{http://dx.doi.org/10.1016/0370-1573(84)90008-5}{\emph{Phys. Rept.} {\bf
  110} (1984) 1--162}.

\bibitem{Haber:1984rc}
H.~E. Haber and G.~L. Kane, \emph{{The Search for Supersymmetry: Probing
  Physics Beyond the Standard Model}},
  \href{http://dx.doi.org/10.1016/0370-1573(85)90051-1}{\emph{Phys. Rept.} {\bf
  117} (1985) 75--263}.

\bibitem{Borowka:2015ura}
S.~Borowka, T.~Hahn, S.~Heinemeyer, G.~Heinrich and W.~Hollik,
  \emph{{Renormalization scheme dependence of the two-loop QCD corrections to
  the neutral Higgs-boson masses in the MSSM}},
  \href{http://dx.doi.org/10.1140/epjc/s10052-015-3648-6}{\emph{Eur. Phys. J.}
  {\bf C75} (2015) 424}, [\href{http://arxiv.org/abs/1505.03133}{{\tt
  1505.03133}}].

\bibitem{Goodsell:2016udb}
M.~D. Goodsell and F.~Staub, \emph{{The Higgs mass in the CP violating MSSM,
  NMSSM, and beyond}},
  \href{http://dx.doi.org/10.1140/epjc/s10052-016-4495-9}{\emph{Eur. Phys. J.}
  {\bf C77} (2017) 46}, [\href{http://arxiv.org/abs/1604.05335}{{\tt
  1604.05335}}].

\bibitem{Passehr:2017ufr}
S.~Paßehr and G.~Weiglein, \emph{{Two-loop top and bottom Yukawa corrections
  to the Higgs-boson masses in the complex MSSM}},
  \href{http://dx.doi.org/10.1140/epjc/s10052-018-5665-8}{\emph{Eur. Phys. J.}
  {\bf C78} (2018) 222}, [\href{http://arxiv.org/abs/1705.07909}{{\tt
  1705.07909}}].

\bibitem{Harlander:2017kuc}
R.~V. Harlander, J.~Klappert and A.~Voigt, \emph{{Higgs mass prediction in the
  MSSM at three-loop level in a pure $\overline{{\text {DR}}}$ context}},
  \href{http://dx.doi.org/10.1140/epjc/s10052-017-5368-6}{\emph{Eur. Phys. J.}
  {\bf C77} (2017) 814}, [\href{http://arxiv.org/abs/1708.05720}{{\tt
  1708.05720}}].

\bibitem{Borowka:2018anu}
S.~Borowka, S.~Paßehr and G.~Weiglein, \emph{{Complete two-loop QCD
  contributions to the lightest Higgs-boson mass in the MSSM with complex
  parameters}},
  \href{http://dx.doi.org/10.1140/epjc/s10052-018-6055-y}{\emph{Eur. Phys. J.}
  {\bf C78} (2018) 576}, [\href{http://arxiv.org/abs/1802.09886}{{\tt
  1802.09886}}].

\bibitem{Vega:2015fna}
J.~P. Vega and G.~Villadoro, \emph{{SusyHD: Higgs mass determination in
  supersymmetry}}, \href{http://dx.doi.org/10.1007/JHEP07(2015)159}{\emph{JHEP}
  {\bf 07} (2015) 159}, [\href{http://arxiv.org/abs/1504.05200}{{\tt
  1504.05200}}].

\bibitem{Lee:2015uza}
G.~Lee and C.~E.~M. Wagner, \emph{{Higgs bosons in heavy supersymmetry with an
  intermediate m$_A$}},
  \href{http://dx.doi.org/10.1103/PhysRevD.92.075032}{\emph{Phys. Rev.} {\bf
  D92} (2015) 075032}, [\href{http://arxiv.org/abs/1508.00576}{{\tt
  1508.00576}}].

\bibitem{Bagnaschi:2017xid}
E.~Bagnaschi, J.~Pardo~Vega and P.~Slavich, \emph{{Improved determination of
  the Higgs mass in the MSSM with heavy superpartners}},
  \href{http://dx.doi.org/10.1140/epjc/s10052-017-4885-7}{\emph{Eur. Phys. J.}
  {\bf C77} (2017) 334}, [\href{http://arxiv.org/abs/1703.08166}{{\tt
  1703.08166}}].

\bibitem{Bahl:2018jom}
H.~Bahl and W.~Hollik, \emph{{Precise prediction of the MSSM Higgs boson masses
  for low $M_{A}$}},
  \href{http://dx.doi.org/10.1007/JHEP07(2018)182}{\emph{JHEP} {\bf 07} (2018)
  182}, [\href{http://arxiv.org/abs/1805.00867}{{\tt 1805.00867}}].

\bibitem{Harlander:2018yhj}
R.~V. Harlander, J.~Klappert, A.~D. Ochoa~Franco and A.~Voigt, \emph{{The light
  CP-even MSSM Higgs mass resummed to fourth logarithmic order}},
  \href{http://dx.doi.org/10.1140/epjc/s10052-018-6351-6}{\emph{Eur. Phys. J.}
  {\bf C78} (2018) 874}, [\href{http://arxiv.org/abs/1807.03509}{{\tt
  1807.03509}}].

\bibitem{Hahn:2013ria}
T.~Hahn, S.~Heinemeyer, W.~Hollik, H.~Rzehak and G.~Weiglein,
  \emph{{High-precision predictions for the light CP-even Higgs boson mass of
  the Minimal Supersymmetric Standard Model}},
  \href{http://dx.doi.org/10.1103/PhysRevLett.112.141801}{\emph{Phys. Rev.
  Lett.} {\bf 112} (2014) 141801}, [\href{http://arxiv.org/abs/1312.4937}{{\tt
  1312.4937}}].

\bibitem{Bahl:2016brp}
H.~Bahl and W.~Hollik, \emph{{Precise prediction for the light MSSM Higgs boson
  mass combining effective field theory and fixed-order calculations}},
  \href{http://dx.doi.org/10.1140/epjc/s10052-016-4354-8}{\emph{Eur. Phys. J.}
  {\bf C76} (2016) 499}, [\href{http://arxiv.org/abs/1608.01880}{{\tt
  1608.01880}}].

\bibitem{Athron:2016fuq}
P.~Athron, J.~Park, T.~Steudtner, D.~St{\"o}ckinger and A.~Voigt,
  \emph{{Precise Higgs mass calculations in (non-)minimal supersymmetry at both
  high and low scales}},
  \href{http://dx.doi.org/10.1007/JHEP01(2017)079}{\emph{JHEP} {\bf 01} (2017)
  079}, [\href{http://arxiv.org/abs/1609.00371}{{\tt 1609.00371}}].

\bibitem{Staub:2017jnp}
F.~Staub and W.~Porod, \emph{{Improved predictions for intermediate and heavy
  Supersymmetry in the MSSM and beyond}},
  \href{http://dx.doi.org/10.1140/epjc/s10052-017-4893-7}{\emph{Eur. Phys. J.}
  {\bf C77} (2017) 338}, [\href{http://arxiv.org/abs/1703.03267}{{\tt
  1703.03267}}].

\bibitem{Bahl:2017aev}
H.~Bahl, S.~Heinemeyer, W.~Hollik and G.~Weiglein, \emph{{Reconciling EFT and
  hybrid calculations of the light MSSM Higgs-boson mass}},
  \href{http://dx.doi.org/10.1140/epjc/s10052-018-5544-3}{\emph{Eur. Phys. J.}
  {\bf C78} (2018) 57}, [\href{http://arxiv.org/abs/1706.00346}{{\tt
  1706.00346}}].

\bibitem{Athron:2017fvs}
P.~Athron, M.~Bach, D.~Harries, T.~Kwasnitza, J.~Park, D.~Stöckinger et~al.,
  \emph{{FlexibleSUSY 2.0: Extensions to investigate the phenomenology of SUSY
  and non-SUSY models}},
  \href{http://dx.doi.org/10.1016/j.cpc.2018.04.016}{\emph{Comput. Phys.
  Commun.} {\bf 230} (2018) 145--217},
  [\href{http://arxiv.org/abs/1710.03760}{{\tt 1710.03760}}].

\bibitem{Heinemeyer:1998yj}
S.~Heinemeyer, W.~Hollik and G.~Weiglein, \emph{{FeynHiggs: A Program for the
  calculation of the masses of the neutral CP even Higgs bosons in the MSSM}},
  \href{http://dx.doi.org/10.1016/S0010-4655(99)00364-1}{\emph{Comput. Phys.
  Commun.} {\bf 124} (2000) 76--89},
  [\href{http://arxiv.org/abs/hep-ph/9812320}{{\tt hep-ph/9812320}}].

\bibitem{Heinemeyer:1998np}
S.~Heinemeyer, W.~Hollik and G.~Weiglein, \emph{{The masses of the neutral
  CP-even Higgs bosons in the MSSM: Accurate analysis at the two loop level}},
  \href{http://dx.doi.org/10.1007/s100529900006,
  10.1007/s100520050537}{\emph{Eur. Phys. J.} {\bf C9} (1999) 343--366},
  [\href{http://arxiv.org/abs/hep-ph/9812472}{{\tt hep-ph/9812472}}].

\bibitem{Hahn:2009zz}
T.~Hahn, S.~Heinemeyer, W.~Hollik, H.~Rzehak and G.~Weiglein, \emph{{FeynHiggs:
  A program for the calculation of MSSM Higgs-boson observables - Version
  2.6.5}}, \href{http://dx.doi.org/10.1016/j.cpc.2009.02.014}{\emph{Comput.
  Phys. Commun.} {\bf 180} (2009) 1426--1427}.

\bibitem{Degrassi:2002fi}
G.~Degrassi, S.~Heinemeyer, W.~Hollik, P.~Slavich and G.~Weiglein,
  \emph{{Towards high precision predictions for the MSSM Higgs sector}},
  \href{http://dx.doi.org/10.1140/epjc/s2003-01152-2}{\emph{Eur. Phys. J.} {\bf
  C28} (2003) 133--143}, [\href{http://arxiv.org/abs/hep-ph/0212020}{{\tt
  hep-ph/0212020}}].

\bibitem{Frank:2006yh}
M.~Frank, T.~Hahn, S.~Heinemeyer, W.~Hollik, H.~Rzehak and G.~Weiglein,
  \emph{{The Higgs boson masses and mixings of the complex MSSM in the
  Feynman-diagrammatic approach}},
  \href{http://dx.doi.org/10.1088/1126-6708/2007/02/047}{\emph{JHEP} {\bf 02}
  (2007) 047}, [\href{http://arxiv.org/abs/hep-ph/0611326}{{\tt
  hep-ph/0611326}}].

\bibitem{Bahl:2018qog}
H.~Bahl, T.~Hahn, S.~Heinemeyer, W.~Hollik, S.~Paßehr, H.~Rzehak et~al.,
  \emph{{Precision calculations in the MSSM Higgs-boson sector with FeynHiggs
  2.14}},  \href{http://arxiv.org/abs/1811.09073}{{\tt 1811.09073}}.

\bibitem{Bahl:2018zmf}
H.~Bahl, E.~Fuchs, T.~Hahn, S.~Heinemeyer, S.~Liebler, S.~Patel et~al.,
  \emph{{MSSM Higgs Boson Searches at the LHC: Benchmark Scenarios for Run 2
  and Beyond}},  \href{http://arxiv.org/abs/1808.07542}{{\tt 1808.07542}}.

\bibitem{Chankowski:1992er}
P.~H. Chankowski, S.~Pokorski and J.~Rosiek, \emph{{Complete on-shell
  renormalization scheme for the minimal supersymmetric Higgs sector}},
  \href{http://dx.doi.org/10.1016/0550-3213(94)90141-4}{\emph{Nucl. Phys.} {\bf
  B423} (1994) 437--496}, [\href{http://arxiv.org/abs/hep-ph/9303309}{{\tt
  hep-ph/9303309}}].

\bibitem{Dabelstein:1994hb}
A.~Dabelstein, \emph{{The one loop renormalization of the MSSM Higgs sector and
  its application to the neutral scalar Higgs masses}},
  \href{http://dx.doi.org/10.1007/BF01624592}{\emph{Z. Phys.} {\bf C67} (1995)
  495--512}, [\href{http://arxiv.org/abs/hep-ph/9409375}{{\tt
  hep-ph/9409375}}].

\bibitem{Pierce:1996zz}
D.~M. Pierce, J.~A. Bagger, K.~T. Matchev and R.-J. Zhang, \emph{{Precision
  corrections in the minimal supersymmetric standard model}},
  \href{http://dx.doi.org/10.1016/S0550-3213(96)00683-9}{\emph{Nucl. Phys.}
  {\bf B491} (1997) 3--67}, [\href{http://arxiv.org/abs/hep-ph/9606211}{{\tt
  hep-ph/9606211}}].

\bibitem{Degrassi:2001yf}
G.~Degrassi, P.~Slavich and F.~Zwirner, \emph{{On the neutral Higgs boson
  masses in the MSSM for arbitrary stop mixing}},
  \href{http://dx.doi.org/10.1016/S0550-3213(01)00343-1}{\emph{Nucl. Phys.}
  {\bf B611} (2001) 403--422}, [\href{http://arxiv.org/abs/hep-ph/0105096}{{\tt
  hep-ph/0105096}}].

\bibitem{Brignole:2001jy}
A.~Brignole, G.~Degrassi, P.~Slavich and F.~Zwirner, \emph{{On the
  $O(\alpha_t^2$) two loop corrections to the neutral Higgs boson masses in the
  MSSM}}, \href{http://dx.doi.org/10.1016/S0550-3213(02)00184-0}{\emph{Nucl.
  Phys.} {\bf B631} (2002) 195--218},
  [\href{http://arxiv.org/abs/hep-ph/0112177}{{\tt hep-ph/0112177}}].

\bibitem{Brignole:2002bz}
A.~Brignole, G.~Degrassi, P.~Slavich and F.~Zwirner, \emph{{On the two loop
  sbottom corrections to the neutral Higgs boson masses in the MSSM}},
  \href{http://dx.doi.org/10.1016/S0550-3213(02)00748-4}{\emph{Nucl. Phys.}
  {\bf B643} (2002) 79--92}, [\href{http://arxiv.org/abs/hep-ph/0206101}{{\tt
  hep-ph/0206101}}].

\bibitem{Dedes:2003km}
A.~Dedes, G.~Degrassi and P.~Slavich, \emph{{On the two loop Yukawa corrections
  to the MSSM Higgs boson masses at large $\tan\beta$}},
  \href{http://dx.doi.org/10.1016/j.nuclphysb.2003.08.033}{\emph{Nucl. Phys.}
  {\bf B672} (2003) 144--162}, [\href{http://arxiv.org/abs/hep-ph/0305127}{{\tt
  hep-ph/0305127}}].

\bibitem{Heinemeyer:2004xw}
S.~Heinemeyer, W.~Hollik, H.~Rzehak and G.~Weiglein, \emph{{High-precision
  predictions for the MSSM Higgs sector at $O(\alpha_b \alpha_s)$}},
  \href{http://dx.doi.org/10.1140/epjc/s2005-02112-6}{\emph{Eur. Phys. J.} {\bf
  C39} (2005) 465--481}, [\href{http://arxiv.org/abs/hep-ph/0411114}{{\tt
  hep-ph/0411114}}].

\bibitem{Heinemeyer:2007aq}
S.~Heinemeyer, W.~Hollik, H.~Rzehak and G.~Weiglein, \emph{{The Higgs sector of
  the complex MSSM at two-loop order: QCD contributions}},
  \href{http://dx.doi.org/10.1016/j.physletb.2007.07.030}{\emph{Phys. Lett.}
  {\bf B652} (2007) 300--309}, [\href{http://arxiv.org/abs/0705.0746}{{\tt
  0705.0746}}].

\bibitem{Hollik:2014wea}
W.~Hollik and S.~Paßehr, \emph{{Two-loop top-Yukawa-coupling corrections to
  the Higgs boson masses in the complex MSSM}},
  \href{http://dx.doi.org/10.1016/j.physletb.2014.04.026}{\emph{Phys. Lett.}
  {\bf B733} (2014) 144--150}, [\href{http://arxiv.org/abs/1401.8275}{{\tt
  1401.8275}}].

\bibitem{Hollik:2014bua}
W.~Hollik and S.~Paßehr, \emph{{Higgs boson masses and mixings in the complex
  MSSM with two-loop top-Yukawa-coupling corrections}},
  \href{http://dx.doi.org/10.1007/JHEP10(2014)171}{\emph{JHEP} {\bf 10} (2014)
  171}, [\href{http://arxiv.org/abs/1409.1687}{{\tt 1409.1687}}].

\bibitem{Hollik:2015ema}
W.~Hollik and S.~Paßehr, \emph{{Two-loop top-Yukawa-coupling corrections to
  the charged Higgs-boson mass in the MSSM}},
  \href{http://dx.doi.org/10.1140/epjc/s10052-015-3558-7}{\emph{Eur. Phys. J.}
  {\bf C75} (2015) 336}, [\href{http://arxiv.org/abs/1502.02394}{{\tt
  1502.02394}}].

\bibitem{Hahn:2015gaa}
T.~Hahn and S.~Paßehr, \emph{{Implementation of the
  $\mathcal{O}{\left(\alpha_t^2\right)}$ MSSM Higgs-mass corrections in
  $\tt{FeynHiggs}$}},
  \href{http://dx.doi.org/10.1016/j.cpc.2017.01.026}{\emph{Comput. Phys.
  Commun.} {\bf 214} (2017) 91--97},
  [\href{http://arxiv.org/abs/1508.00562}{{\tt 1508.00562}}].

\bibitem{Borowka:2014wla}
S.~Borowka, T.~Hahn, S.~Heinemeyer, G.~Heinrich and W.~Hollik,
  \emph{{Momentum-dependent two-loop QCD corrections to the neutral Higgs-boson
  masses in the MSSM}},
  \href{http://dx.doi.org/10.1140/epjc/s10052-014-2994-0}{\emph{Eur. Phys. J.}
  {\bf C74} (2014) 2994}, [\href{http://arxiv.org/abs/1404.7074}{{\tt
  1404.7074}}].

\bibitem{Sperling:2013xqa}
M.~Sperling, D.~St{\"o}ckinger and A.~Voigt, \emph{{Renormalization of vacuum
  expectation values in spontaneously broken gauge theories: Two-loop
  results}}, \href{http://dx.doi.org/10.1007/JHEP01(2014)068}{\emph{JHEP} {\bf
  01} (2014) 068}, [\href{http://arxiv.org/abs/1310.7629}{{\tt 1310.7629}}].

\bibitem{Appelquist:1974tg}
T.~Appelquist and J.~Carazzone, \emph{{Infrared Singularities and Massive
  Fields}}, \href{http://dx.doi.org/10.1103/PhysRevD.11.2856}{\emph{Phys. Rev.}
  {\bf D11} (1975) 2856}.

\bibitem{Collins:1984xc}
J.~C. Collins, \emph{{Renormalization}}, vol.~26 of \emph{Cambridge Monographs
  on Mathematical Physics}.
\newblock Cambridge University Press, Cambridge, 1986,
  \href{http://dx.doi.org/10.1017/CBO9780511622656}{10.1017/CBO9780511622656}.

\bibitem{Tkachov:1991ev}
F.~V. Tkachov, \emph{{Euclidean asymptotic expansions of Green functions of
  quantum fields. 1. Expansions of products of singular functions}},
  \href{http://dx.doi.org/10.1142/S0217751X93000850}{\emph{Int. J. Mod. Phys.}
  {\bf A8} (1993) 2047--2117}, [\href{http://arxiv.org/abs/hep-ph/9612284}{{\tt
  hep-ph/9612284}}].

\bibitem{Pivovarov:1991du}
G.~B. Pivovarov and F.~V. Tkachov, \emph{{Euclidean asymptotic expansions of
  Green functions of quantum fields. 2. Combinatorics of the AS operation}},
  \href{http://dx.doi.org/10.1142/S0217751X93000898}{\emph{Int. J. Mod. Phys.}
  {\bf A8} (1993) 2241--2286}, [\href{http://arxiv.org/abs/hep-ph/9612287}{{\tt
  hep-ph/9612287}}].

\bibitem{Smirnov:1990rz}
V.~A. Smirnov, \emph{{Asymptotic expansions in limits of large momenta and
  masses}}, \href{http://dx.doi.org/10.1007/BF02102092}{\emph{Commun. Math.
  Phys.} {\bf 134} (1990) 109--137}.

\bibitem{Smirnov:1994tg}
V.~A. Smirnov, \emph{{Asymptotic expansions in momenta and masses and
  calculation of Feynman diagrams}},
  \href{http://dx.doi.org/10.1142/S0217732395001617}{\emph{Mod. Phys. Lett.}
  {\bf A10} (1995) 1485--1500},
  [\href{http://arxiv.org/abs/hep-th/9412063}{{\tt hep-th/9412063}}].

\bibitem{Smirnov:1996ng}
V.~A. Smirnov, \emph{{Asymptotic expansions of Feynman diagrams on the mass
  shell in momenta and masses}},
  \href{http://dx.doi.org/10.1016/S0370-2693(96)01697-8}{\emph{Phys. Lett.}
  {\bf B394} (1997) 205--210}, [\href{http://arxiv.org/abs/hep-th/9608151}{{\tt
  hep-th/9608151}}].

\bibitem{Frank:2002qf}
M.~Frank, S.~Heinemeyer, W.~Hollik and G.~Weiglein, \emph{{FeynHiggs1.2: Hybrid
  MS-bar / on-shell renormalization for the CP even Higgs boson sector in the
  MSSM}},  \href{http://arxiv.org/abs/hep-ph/0202166}{{\tt hep-ph/0202166}}.

\bibitem{Heinemeyer:2001iy}
S.~Heinemeyer, W.~Hollik, J.~Rosiek and G.~Weiglein, \emph{{Neutral MSSM Higgs
  boson production at $e+$ $e-$ colliders in the Feynman diagrammatic
  approach}}, \href{http://dx.doi.org/10.1007/s100520100631}{\emph{Eur. Phys.
  J.} {\bf C19} (2001) 535--546},
  [\href{http://arxiv.org/abs/hep-ph/0102081}{{\tt hep-ph/0102081}}].

\bibitem{Fuchs:2016swt}
E.~Fuchs and G.~Weiglein, \emph{{Breit-Wigner approximation for propagators of
  mixed unstable states}},
  \href{http://dx.doi.org/10.1007/JHEP09(2017)079}{\emph{JHEP} {\bf 09} (2017)
  079}, [\href{http://arxiv.org/abs/1610.06193}{{\tt 1610.06193}}].

\bibitem{Fuchs:2017wkq}
E.~Fuchs and G.~Weiglein, \emph{{Impact of CP-violating interference effects on
  MSSM Higgs searches}},
  \href{http://dx.doi.org/10.1140/epjc/s10052-018-5543-4}{\emph{Eur. Phys. J.}
  {\bf C78} (2018) 87}, [\href{http://arxiv.org/abs/1705.05757}{{\tt
  1705.05757}}].

\bibitem{Domingo:2017rhb}
F.~Domingo, P.~Drechsel and S.~Paßehr, \emph{{On-Shell neutral Higgs bosons in
  the NMSSM with complex parameters}},
  \href{http://dx.doi.org/10.1140/epjc/s10052-017-5104-2}{\emph{Eur. Phys. J.}
  {\bf C77} (2017) 562}, [\href{http://arxiv.org/abs/1706.00437}{{\tt
  1706.00437}}].

\end{thebibliography}\endgroup

\end{document}